# Oxysulfide Perovskites: Reduction of the Electronic Band Gap of RbTaO$_3$ by Sulfur Substitution to Enhance Prospective Solar Cell and Thermoelectric Performances

H. Akter[1], M. A. Ali[1,*], M. M. Hossain[1], M. M. Uddin[1], S. H. Naqib[1,2]

[1]Advanced Computational Materials Research Laboratory, Department of Physics, Chittagong University of Engineering and Technology (CUET), Chattogram-4349, Bangladesh

[2]Department of Physics, University of Rajshahi, Rajshahi-6205, Bangladesh

**Abstract**

In this study, the effects of sulfur substitution on the structural, mechanical, electronic, optical, and thermodynamic properties of RbTaO$_{3-x}$S$_x$ have been investigated using the WIEN2k code in the framework of density functional theory (DFT). The cubic phase of RbTaO$_3$ transforms to tetragonal for RbTaO$_2$S and RbTaOS$_2$, the later transforms again to a cubic phase with added sulfur for RbTaS$_3$. The results showed that substituting S for O anions in RbTaO$_3$ effectively decreased the band gap from 2.717 eV to 1.438 eV, 0.286 eV, and 0.103 eV for the RbTaO$_3$, RbTaO$_2$S, RbTaOS$_2$, and RbTaS$_3$ compounds, respectively. The optical constants such as dielectric constants, refractive index, absorption coefficient, photoconductivity, reflectivity and loss function have been calculated and analyzed. The elastic constants and moduli, and their anisotropic nature were also investigated. Finally, the Debye temperature, thermal conductivity, melting temperature, specific capacities and thermal expansion coefficients were computed and analyzed using established formalisms. The reduced band gap (1.438 eV) and high absorption coefficient (~10$^6$ cm$^{-1}$) of RbTaO$_2$S makes it suitable for solar cell applications and for other visible light devices. Reduction of the band gap and phonon thermal conductivity owing to *S*-substitution is expected to enhance thermoelectric performances of the *S*-containing phases.

Corresponding Author: ashrafphy31@cuet.ac.bd

## 1. Introduction

The formula $ABX_3$ represents a large family of materials known as perovskite. The structure exhibits a wide range of flexibility; more than 90% of the metal from the periodic table can be adapted within the structure, which makes them more versatile both from the scientific and industrial points of view [1,2]. X is the anionic sub-lattice in the generic formula $ABX_3$, where X atom can be chosen from the family of halides, chalcogenides, or pnictide (X = F, Cl, Br, I; O, S, Se; N). In addition, mixing of atoms at the A, B, and X sites makes the perovskite family an amazing collection of compounds. The diversity of these compositions and ordering result in a prospective combination of electronic, optical, and magnetic properties, making them suitable for use in electronic devices, sensors, energy storage, and as catalysts [3–6].

Several oxide Perovskites, including $NaTaO_3$ [7], $KTaO_3$ [8], $BaZrO_3$ [9], $SrTiO_3$ [10], $BaTiO_3$ [11], $CaTiO_3$ [12], and $LaFeO_3$ [13], $EuAlO_3$[14] and $YMnO_3$ [15] have demonstrated favorable optical, electrical, structural, and thermal properties. These Perovskite materials offer diverse properties that make them valuable in various fields, including energy harvesting, electronics, catalysis, sensing, and information storage. Alkaline oxide perovskites like $KTaO_3$, $RbTaO_3$, and $CsTaO_3$ semiconductors are also intriguing compounds [16]. Due to their excellent thermal characteristics and electrical band gaps in the visible region, these materials are well-suited for optoelectronic and photovoltaic applications. Alkaline transition metal oxides are now gaining much attention for scientific and industrial applications [17, 18]. $RbTaO_3$, a newly discovered Perovskite oxide, has been proven to exist in a stable cubic phase [19], which is also stable in orthorhombic and tetragonal structures [20]. Recent research on the magnetic and thermoelectric properties of tantalum-based cubic perovskite oxides $RbTaO_3$ has shown that it is a potential candidate for thermoelectric and other related device applications [21]. This intriguing ternary perovskite oxide, $RbTaO_3$, was first explored by Smolenskii and Kozhevnikova [22] and then investigated by Megaw [23].

The abovementioned materials have been thoroughly explored using the first-principles calculations with several approximations, including GGA, LDA, and GGA+U, from which it was noticed that their band gap are around 3 eV, and most of the compounds exhibit notable activity in the UV range. As a sustainable, eco-friendly, and affordable light source, this is crucial for utilizing solar radiation for energy harvesting applications [24]. In light of the limited utilization of solar irradiation by RbTaO$_3$ due to its wide band gap (2.7 eV) [25] and small UV fraction (about 20-30%), significant efforts have to be devoted to enhance its solar radiation dependent applications by narrowing the pristine band gap. One promising approach involves the incorporation of suitable anions at O site of oxide Perovskite like, NaTaO$_3$, RbTaO$_3$ to narrow its band gap and improve its utilization of solar energy [26]. While traditional cation dopants tend to reduce the conduction band-edge potential, anion dopants at the O-sites offer a different avenue. Among various nonmetallic dopants, sulfur (S) stands out as a particularly promising candidate due to its similar valence electron configuration as oxygen. By substituting oxygen sites with sulfur, RbTaO$_3$ can benefit from a shifted valence band edge, reducing the band gap by approximately 1.4 eV without introducing undesirable holes or defects. This approach aligns with recent theoretical findings and parallels the success of S-doped TiO$_2$, preserving the ability to split water into H$_2$ and O$_2$ [27-29]. In fact, S-doped RbTaO$_3$ exhibited negative minimum formation energy, demonstrating its superior thermodynamic stability compared to other anion doped systems. As a result, S anion doping presents a promising avenue to endow RbTaO$_3$ with improved utilization of solar energy.

Prior research by Lu et al. [30] on Sulfur (S) doped SrTiO$_3$ revealed that the non-metal element successfully lowered the band gap from 3.5 eV to 1.44 eV. Another wide-band insulator with a gap of about 3.2 eV is BaTiO$_3$. Using chemical replacements of oxygen vacancies to dope conduction electrons into Ti-3$d$ band lowered their band gap, and it can be converted from a semiconductor to a metal [31,32].

RbTaO$_3$ has a band gap of 2.71 eV, which is higher than that required for solar energy harvesting technology, exceeding the crucial value (1.43 eV) that is normally required to achieve maximum efficiency for solar cells [33]. The band gap of RbTaO$_3$ can be reduced by altering the material's composition, which then changes the features of its electronic structure. Optoelectronic materials can shift to higher energy levels and supply enough energy to start a chemical reaction by absorbing light. By doping or exchanging oxygen with a chalcogenide atom, such as S or Se, visible light activity can be increased [34]. The reduction of the band gap in oxide perovskites through the substitution of S in place of O has been reported [35-40]. Now, a question may arise; why does the band gap reduce by S substitution? [38-40]. By introducing different sulfur (S) contents, the wide band gap of BaZrO$_3$ has been effectively reduced to a range of 1.73 eV to 2.87 eV [38]. The substitution of sulfur (S) in NaTaO$_3$ has resulted in a significant reduction in its band gap, leading to a noteworthy enhancement of the photo-catalytic activity [39]. S doping has successfully modified the band gap of NaTaO$_3$, resulting in a remarkable red shift of the adsorption edge observed through UV-vis diffuse reflectance spectroscopy (DRS) and photocatalytic tests. This effect is attributed to the close alignment of the S-3$p$ orbital energy with the top of the valence bands, surpassing the energy of O-2$p$ orbitals. Upon substituting O with S, the interaction between the 3$p$ orbitals of S and Ta-5$d$ orbitals gives rise to the formation of discrete mid-gap bands positioned just above the valence band. Confirmation of these states was obtained through the observation of splitting slopes near the adsorption threshold, resulting in the extension of an 'add-on tail' into the visible light region. This phenomenon suggests that the valence band of NaTaO$_3$ was broadened due to S anion doping, thereby narrowing the band gap and enhancing the visible-light activity of S-doped NaTaO$_3$ [40]. A recent study by Bennett et al. [41] utilized density functional theory (DFT) method to investigate the reduction of the band gap in BaZrO$_3$ through S doping. The modification of electronic properties in oxide Perovskites through S doping was also explored by Ran et al. [42]. Peng et al. [43] reported on the electronic and defective engineering of electrospun CaMnO$_3$ nanotubes. In a separate study, Zhai et al. [44] utilized a function-confined machine learning method to predict the formation of fractionally doped Perovskite oxides. The investigation conducted by Pilania et al. [45] using DFT method focused on studying the anion order in oxysulfide perovskites and its origins and implications. These recent reports serve as strong motivation for our current study, as they inspire us to select the titled compounds for further investigation.

The main target of our study is to reduce the band gap of the previously synthesized Perovskite compound RbTaO$_3$. Again, reiterating the question, why do we want to reduce the band gap? We have focused on two possible applications of the S-containing perovskites: (i) to make it suitable for use in solar cell application and (ii) is to make the chalcogenide Perovskites suitable for use in thermoelectric devices. If we see the legendary work of W. Shockley and H. J. Queisser [33] to achieve the highest efficiency for solar cell materials, we find that the highest efficiency might be achieved for a material with band gap of about 1.43 eV. For thermoelectric materials, if we see the formula for ZT value [the figure of merit] of a thermoelectric material, we found that ZT values ($ZT = \frac{S^2 \sigma T}{k}$, s is the seebeck coefficient, σ is the electrical conductivity, T is the absolute temperature, and *k* is the thermal conductivity) [46] is directly proportional to the conductivity of a semiconducting materials. Narrowing band is one of the possible ways of increasing the electrical conductivity of the semiconducting materials. Keeping in mind the above two prospects, we want to reduce the band gap of the targeted oxide perovskite RbTaO$_3$ via anion doping.

Therefore, we aimed to carry out a DFT-based study to investigate the effect of S substitution on the physical properties of RbTaO$_3$, where S anions will replace the O anions. It is found that the band gap of RbTaO$_2$S is 1.438 eV, which is very close to the S-Q limit [33] and the thermal conductivity of S substituted perovskite is found to be lowered significantly, which is also crucial in association with reducing band gap for enhancing the ZT values of thermoelectric materials. In sections to follow, the physical properties of S-substituted perovskites will be analyzed to reveal their possible applications.

## 2. Computational Methodology

The present investigation validates the first-principles calculations for oxide perovskites by utilizing the full-potential linear augmented plane wave (FP-LAPW) method [47] that is based on the density functional theory (DFT) [48]. The calculations were performed using the WIEN2k code [49]. Optimization of the crystal structure and calculation of mechanical properties were carried out using the Perdew-Burke-Ernzerhof generalized gradient approximation (GGA-PBE) [45,51], while electronic and optical properties were determined using the modified Becke-Johnson potential (TB-mBJ) [52]. A total of 10 × 10 × 10 meshes (1000 K-points) were sampled in the first Brillouin zone of reciprocal space. To ensure convergence, this study considered FP-LAPW basis functions up to $R_{MT} \times K_{max} = 6$ (where $R_{MT}$ is the minimum radius of the muffin-tin spheres) and $K_{max}$ indicating the magnitude of the largest *k*-vector in the plane wave expansions. The energy convergence was set to $10^{-5}$ Ry, and the charge density was Fourier expanded up to $G_{max} = 12$. The density of states was calculated using the tetrahedron approach. Using the IRelast package, different parameters related to elastic properties were computed [53]. Finally, phonon dispersion curves were calculated using the PHONOPY software [54].

## 3. Results and discussion

### *3.1. Structural Properties and Stability*

RbTaO$_3$ is a perovskite with a cubic crystal structure that falls under space group Pm-3m (#221). The atomic positions are: Rb at the center of the unit cell (0, 0, 0), Ta at the corner (0.5, 0.5, 0.5), and O at the face center (0, 0.5, 0.5) [55]. To optimize the unit cell volume and lattice parameters for all phases of the RbTaO$_{3-x}$S$_x$ crystal, we used the atomic positions and applied the generalized gradient approximation (GGA). One or two sulfur atoms substitution in cubic RbTaO$_3$ results in the formation of RbTaO$_2$S and RbTaOS$_2$, respectively. The atomic positions within RbTaO$_2$S are as follows: Rb (0, 0, 0), Ta (0.5, 0.5, 0.5), O (0.5, 0, 0.5), and S (0, 0.5, 0.5).

In the case of RbTaOS$_2$, the atomic positions are as follows: Rb (0, 0, 0), Ta (0.5, 0.5, 0.5), O (0.5, 0.5, 0), and S (0.5, 0, 0.5). This substitution triggers a transformation of the crystal structure from cubic to tetragonal. The space group for these structural transitions is P4/mmm (123).

In addition, substituting three sulfur atoms in the parent compound leads to the formation of RbTaS$_3$, which adopts a cubic structure with the same space group and same atomic position as RbTaO$_3$.

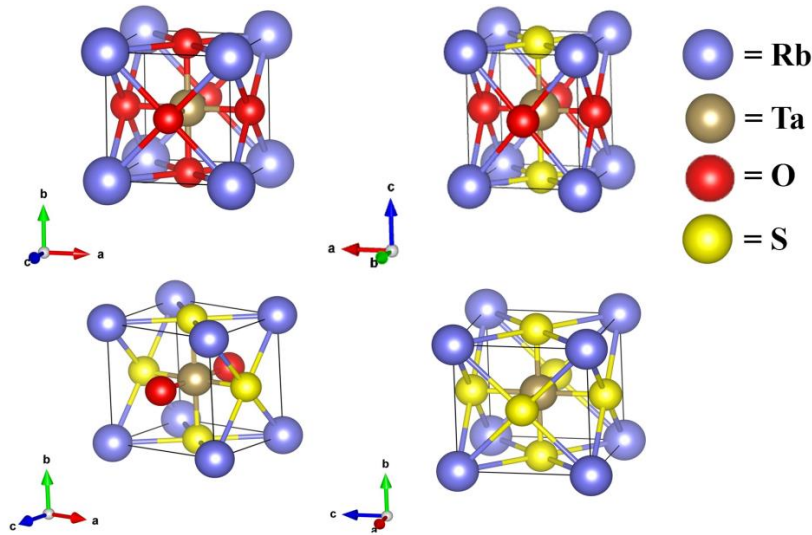

**Fig. 1:** The schematic unit cells of (a) RbTaO$_3$ (b) RbTaO$_2$S (c) RbTaOS$_2$ and (d) RbTaS$_3$

We performed volume optimization for the two cubic structures RbTaO$_3$ and RbTaS$_3$, and got their minimum energy volume shown in Fig. 2(a). The structural stability and fitting of the total energy versus volume for those two cubic compounds have been studied using the third-order Birch-Murnaghan equation of state, which can be described as:

$$E(V) = E_0 + \frac{B_0}{B_0'}\left[\frac{(V/V0)n'}{B_0'-1} + 1\right] - \frac{BV_0}{B'} \qquad (1)$$

The symbol $E_0$ represents the total energy per primitive cell in the ground state, $B$ denotes the bulk modulus, and $V_0$ represents the volume at static equilibrium.

We performed structure optimization in two steps for the other two tetragonal structures. In the initial stage, the volume was varied while maintaining a fixed *c/a* ratio. By performing the Birch-Murnaghan equation of state (EOS) as given in eq. (1) this varied volume was fitted, as shown in Fig. 2(b).

Subsequently, the equilibrium volume was held constant while the *c/a* ratio was varied, which was determined through a parabolic fit. The result of this geometrical optimization, like lattice parameters and other structural data for both cubic and tetragonal compounds, are included in Table 1. The deviation of lattice parameters for Parent compounds is very low (~ 0.1%). Thus, it justifies the present calculation accuracy.

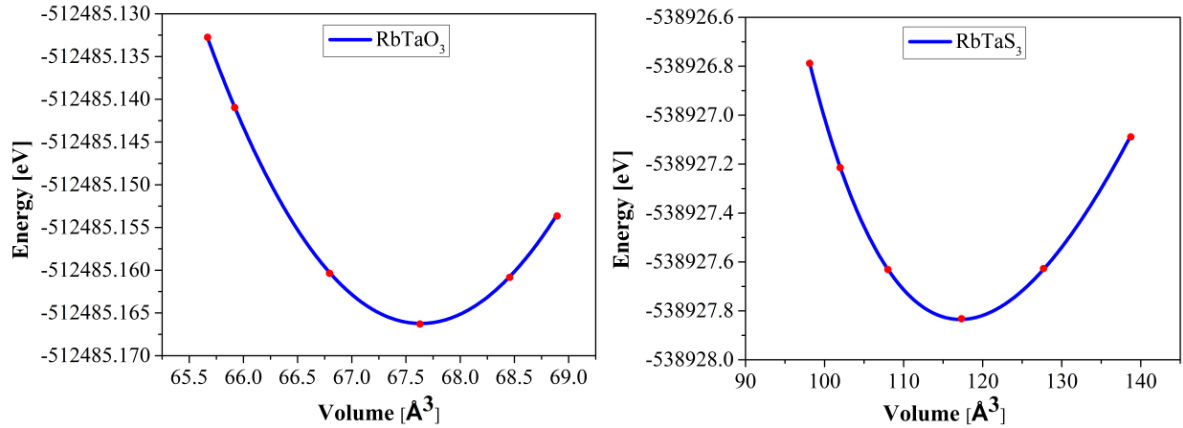

**Fig. 2(a):** Volume Vs energy curve of cubic RbTaO$_3$ and RbTaS$_3$.

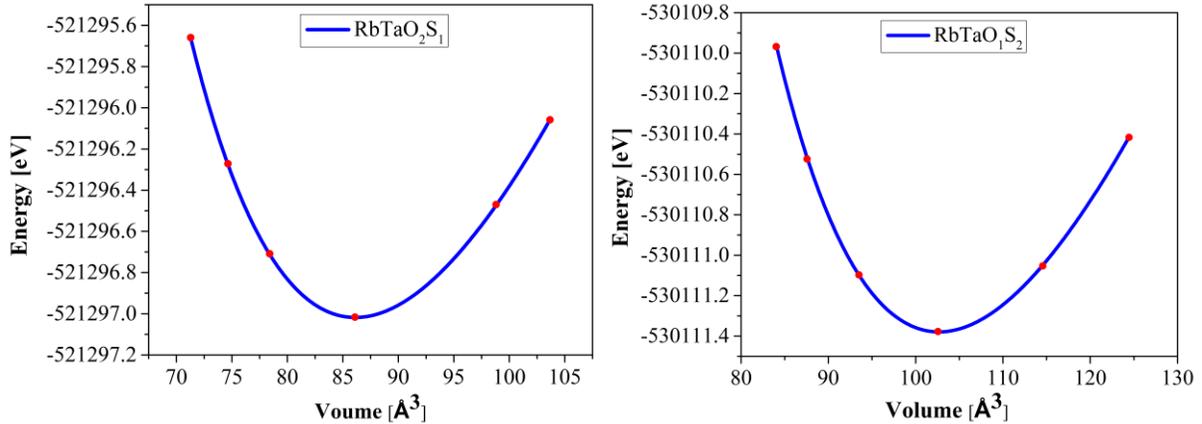

**Fig. 2(b):** Volume Vs energy curve of tetragonal RbTaO$_2$S and RbTaOS$_2$.

The chemical stability of the Perovskite semiconductor RbTaO$_{3-x}$S$_x$ was investigated by calculating the formation energy ($E_f$). The procedure is as follows [57]:

$$E_{for}^{RbTaO_{3-x}S_x} = \frac{E_{TotaL}^{RbTaO_{3-x}S_x} - (AE_{solid}^{Rb} + BE_{solid}^{Ta} + CE_{solid}^{O} + DE_{solid}^{S})}{A+B+C+D} \quad (2)$$

In the above equation, A is the number of atoms of Rb, B is the number of atoms of Ta, and C and D are the numbers of atoms of O and S in the unit cell, respectively. $E_{TotaL}^{RbTaO_{3-x}S_x}$, $E_{solid}^{Rb}$, $E_{solid}^{Ta}$, $E_{solid}^{O}$ and $E_{solid}^{S}$ represent the total energies of the RbTaO$_{3-x}$S$_x$ perovskite semiconductor and the solid forms of Rb, Ta, O, and S in their stable structures, respectively. The predicted formation energies for the compounds RbTaO$_3$, RbTaO$_2$S, RbTaOS$_2$, and RbTaS$_3$ are as follows: -2.8328 eV/atom, -1.6410 eV/atom, -0.9524 eV/atom, and -0.6809 eV/atom, respectively.

**Table 1:** The optimized crystallographic lattice parameters, *a* and *c* (both in Å), volume of unit cell V(Å$^3$) of perovskite RbTaO$_{3-x}$S$_x$ compounds and the percentage of deviation from the reference data.

| Compounds | *a* (Å) | *c* (Å) | % of deviation | Volume (Å$^3$) | Ref |
| --- | --- | --- | --- | --- | --- |
| RbTaO$_3$ | 4.0750 | - | 0.1769 | 67.838 | 4.0678 [56] |
| RbTaO$_2$S | 4.1962 | 4.8936 | - | 86.129 | This study |
| RbTaOS$_2$ | 4.9698 | 4.1587 | - | 102.67 | This study |
| RbTaS$_3$ | 4.8940 | - | - | 117.217 | This study |

**Dynamical stability**

The capacity of a system to tolerate small atomic displacements due to thermal or imposed periodic motions is known as dynamic stability, which can be determined by calculating the phonon dispersions of a material using either a finite displacement method [56] or density functional perturbation theory [59]. In our study, we utilized the finite displacement methodology, which was implemented using the PHONOPY software [60] to derive phonon dispersions. The force constants for the cubic phase were calculated using a supercell with dimensions of (2 × 2 × 2). For the tetragonal phase, calculations were performed using both a (1 × 1 × 1) and a (2 × 2 × 2) supercell. In both cases, atomic displacements of 0.01 Å were applied along the lattice vectors during the calculations. We analyzed the phonon frequency range over the entire Brillouin zone, where a positive frequency indicates stability and a negative frequency signifies instability. Fig. 3 indicates the absence of negative frequency phonon mode in RbTaO$_{3-x}$S$_x$ compounds, demonstrating their dynamic stability. Unfortunately, no experimental investigations are available on the dynamic characteristics these compounds.

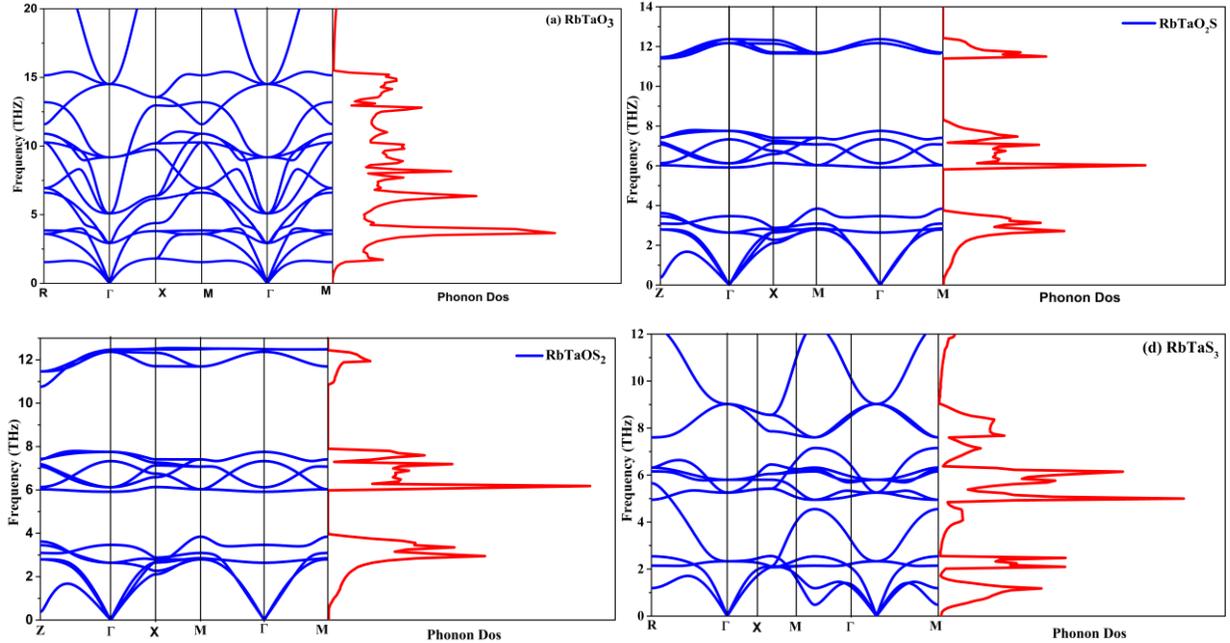

**Fig. 3:** Phonon dispersion curve and phonon DOS of (a) RbTaO$_3$, (b) RbTaO$_2$S, (c) RbTaOS$_2$, and (d) RbTaS$_3$.

The PDCs provide additional information regarding the studied compounds. Due to the presence of five atoms in the unit cell of these compounds, the PDCs exhibit a total of 15 vibrational modes. Three of these modes are acoustic modes, while the remaining 12 are optical modes. At low values of crystal wave vector, the three lowest modes exhibit a dispersion curve in the form of $\omega = vk$, which corresponds to the $\omega(k)$ relationship observed in sound waves and belongs to the acoustic branch. Out-of-phase oscillations of atoms induced by photon excitation give rise to the upper vibrational modes, which constitute the optical branch. At the $\Gamma$-point, the frequency of the acoustic modes is zero, and there is no phononic band gap observed between the acoustic modes and the lower optical branches because of their overlapping.

*3.2 Electronic Properties*

The band structure, electronic density of states (DOS), and charge density distribution of RbTaO$_{3-x}$S$_x$ have been investigated to illustrate the electronic properties and chemical bonding nature of the material. It is reported that the modified Becke-Johnson exchange potential can yield highly accurate energy band gaps of semiconducting and insulating materials [56,61]. Therefore, we have tried searching the band gaps using the TB-mBJ and GGA-PBE approaches. Table 2 displays the estimated band gap of RbTaO$_{3-x}$S$_x$ compounds; it is evident that the bandgap varies significantly for different functionals. The electronic band structure and density of states (DOS) calculated using the TB-mBJ potential are depicted in Fig. 4 and Fig. 5. The band structure results have been presented for RbTaO$_{3-x}$S$_x$ along the high symmetry directions within the Brillouin zone, covering the *R, Γ, X, M, Γ* paths and *Z, Γ, X, M, Γ* for cubic and tetragonal structures, respectively. The Fermi-Level is set at 0 eV. In Fig. 4, it is shown that the top of the valence band (TVB) and the bottom of the conduction band (BCB) are not located at the same k-points, indicating the indirect band gap semiconducting behavior of the studied phases.

**Table 2:** Estimated bandgaps of perovskite RbTaO$_{3-x}$S$_x$ compounds.

| Compounds | Approach | Bandgap (eV) | Ref |
|---|---|---|---|
| RbTaO$_3$ | GGA-PBE | 2.219 | This study |
| | | 2.053 | [56] |
| | | 2.16 | [62] |
| | TB-mBJ | 2.717 | This study |
| | | 2.75 | [25] |
| RbTaO$_2$S$_1$ | GGA-PBE | 0.710 | This study |
| | TB-mBJ | 1.438 | This study |
| RbTaO$_1$S$_2$ | GGA-PBE | 0 | This study |
| | TB-mBJ | 0.286 | This study |
| RbTaS$_3$ | GGA-PBE | 0 | This study |
| | TB-mBJ | 0.103 | This study |

The highest energy part of the valence band (near the Fermi level) mainly occurs due to the *p* states of O or S atoms and the *d* state of Ta atoms of RbTaO$_{3-x}$S$_x$. Similarly, the lowest of the conduction bands is also due to hybridization between the *d* states of the Ta atom. The computed band gap of RbTaO$_3$ using GGA and TB-mBj potential are 2.219 eV and 2.717eV, respectively. These values are reasonably consistent with other predicted band gap values as reported in Table 2. The band gap calculated using TB-mBJ potential for other three compounds are 1.438 eV (RbTaO$_2$S), 0.286 eV (RbTaOS$_2$), and 0.103 eV (RbTaS$_3$), respectively. These results indicate that substitution of S can dramatically decrease the band gap of RbTaO$_3$. The band gap of RbTaO$_2$S is very close to that of Shockley-Queiser limit for band gap of solar cell materials to achieve highest efficiency [27]. Moreover, the band gap is further reduced with increased S contents. Thus, it is expected that electrical conductivity of the S-substituted compositions could be increased with the reduction of band gap.

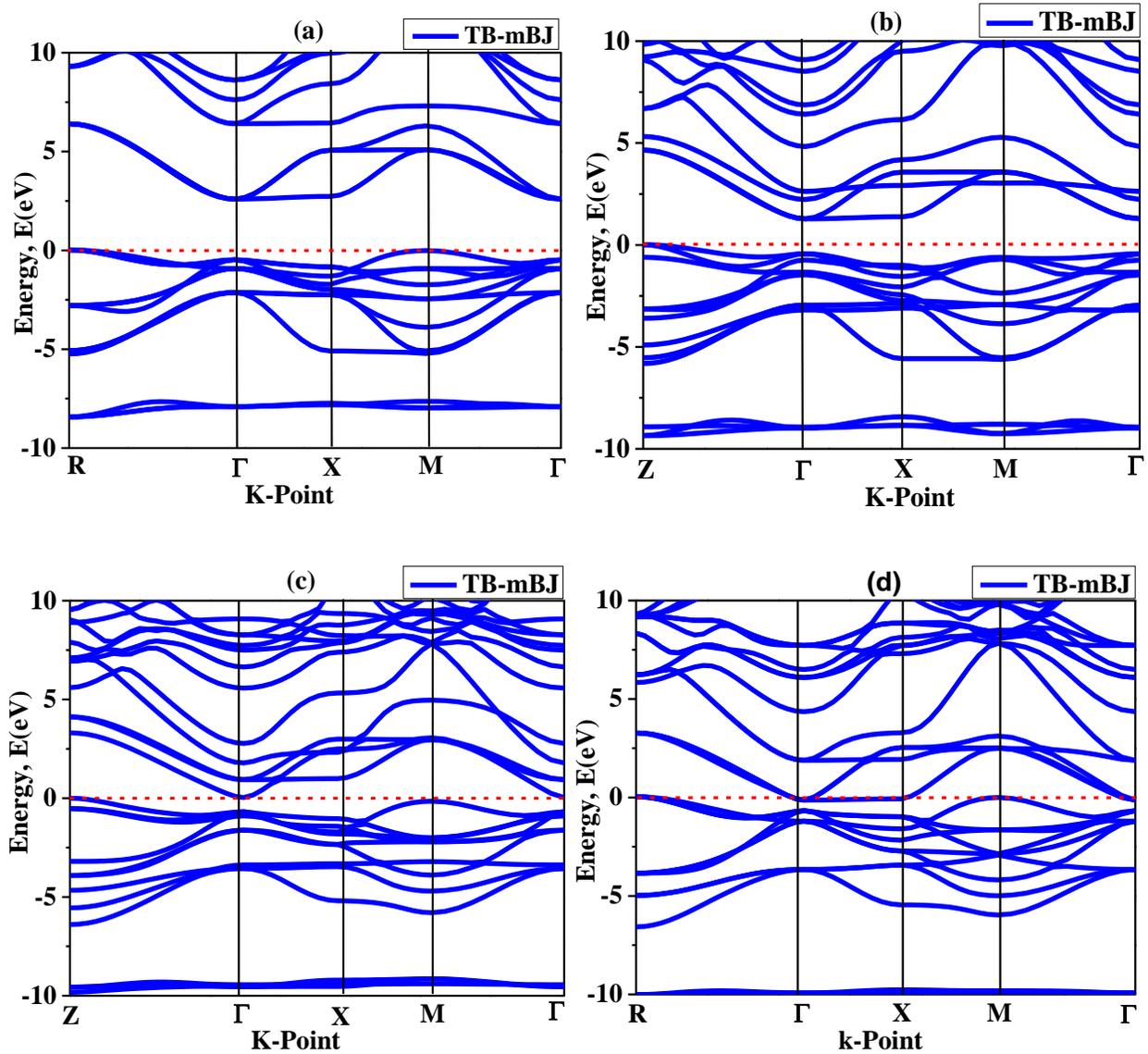

**Fig. 4:** Electronic band structures of a) RbTaO$_3$, b) RbTaO$_2$S, c) RbTaOS$_2$, and d) RbTaS$_3$ compounds.

The thermal conductivity is also found to be decreasing for S-containing phases [section 3.4]. As the ZT values of the TE materials is directly (inversely) proportional to the electrical conductivity (thermal conductivity), thus, an improvement of the thermoelectric properties is expected for the S-substituted compositions. No experimental data regarding these parameters for RbTaO$_3$ is currently available in the literature.

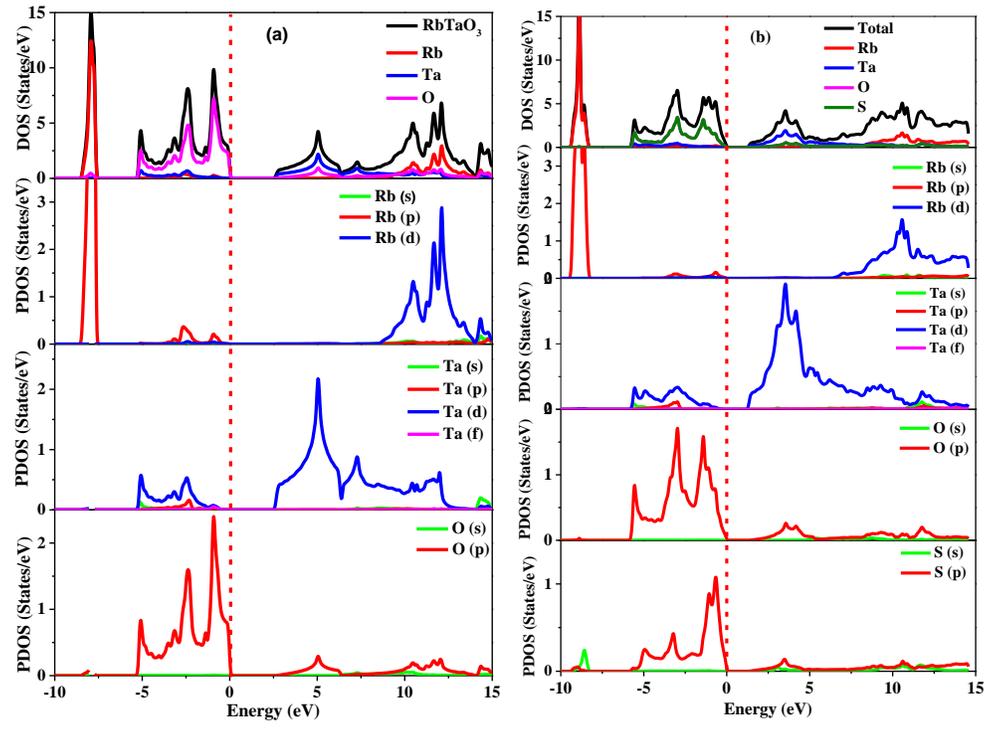
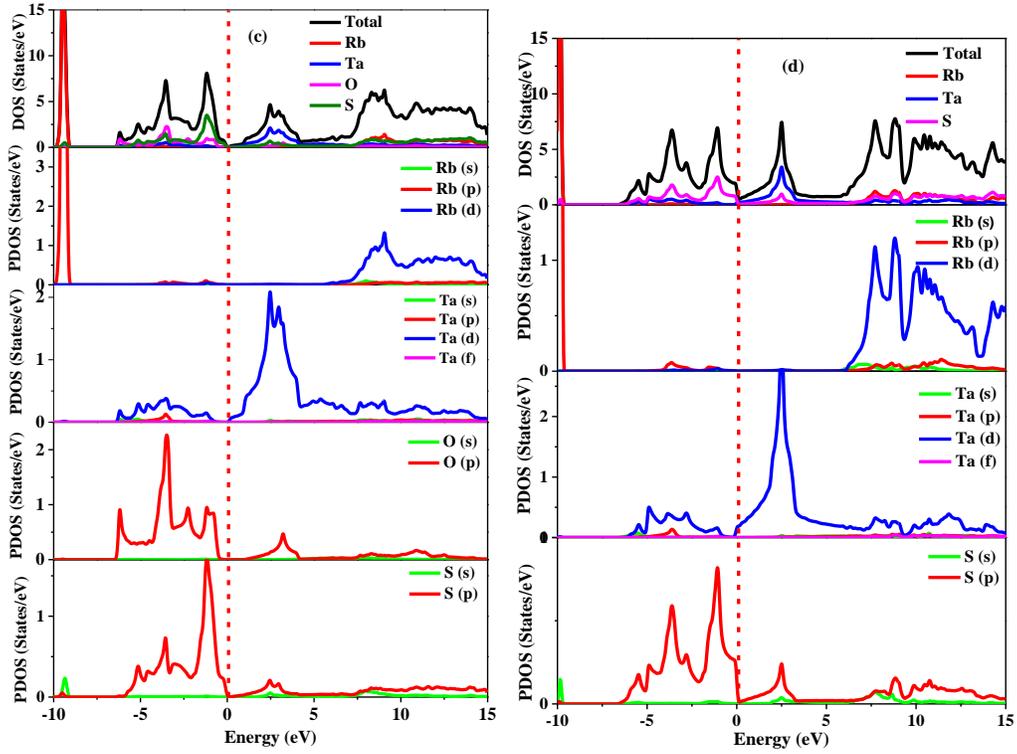

**Fig. 5:** Density of states (DOS) of (a) RbTaO$_3$, (b) RbTaO$_2$S, (c) RbTaOS$_2$ and (d) RbTaS$_3$ compounds.

The total and partial density of states (DOS) has been studied to understand the orbital contributions to atomic bonding and formation of valence and conduction states. The band structure and density of state of various compositions exhibit considerable similarities. The PDOS plot [Fig. 5(a)] for RbTaO$_3$ shows that Rb-$p$, Ta-$d$, and O-$p$ states contribute to the valence band in the deep region, suggesting a strong hybridization between.

Upon substituting one, two, and three sulfur atoms in RbTaO$_3$, it can be observed that the contribution of Rb-$p$ states in the valence band near the Fermi level is significantly reduced. An increase in the contribution of S-$p$ states in the valence band compensates for this reduction in the Rb-$p$ contribution.

Thus, the substitution of S for O in RbTaO$_3$ alters the material's electronic structure, leading to a change in the contributions of the atomic orbitals to the valence band. Specifically, the S-$p$ states start to contribute significantly in the valence band, and the Rb-$p$ contribution decreases.

The conduction band of RbTaO$_3$ exhibits a significant hybridization between Ta-$d$ and O-$p$ states near the Fermi level, primarily responsible for forming the band gap. At higher energy levels in the conduction band, around 10 eV, there is hybridization among Rb-$d$, Ta-$d$, and O-$p$ states, although O-$p$ states make a minor contribution. When one, two, or three sulfur atoms are substituted for oxygen, S-$p$ states start contributing near the Fermi level and hybridize with Ta-$d$ states. Pseudo-states of tantalum (Ta-$d$) play a significant role in forming valence and conduction bands.

It is common to use electronic charge density contour plots to provide accurate explanations of bonding character, as done in the study of the RbTaO$_{3-x}$S$_x$ compounds. These plots are generated by calculating the charge density from a converged wave function using the TB-mBJ scheme, which allows for a comprehensive understanding of bonding properties and charge transfer in the material. Fig. 6 depicts the resulting charge density contour plots for different planes, revealing that the bonds in RbTaO$_{3-x}$S$_x$ are primarily ionic in nature with some covalent contributions.

The spherical charge density contours surrounding the A-cations (Rb) indicate their ionic character. On the other hand, the degree of electron sharing between Ta and O atoms can be inferred from the compaction rate of the isolines around them and the significant overlap of elliptical isolines, which indicates the covalent nature of the Ta-O bonds. In addition to indicating ionic or covalent character, the spherical charge density contours also show the transfer of charges from the valence to the conduction bands. This transfer results in the accumulation or depletion of charges around the corresponding oxygen and tantalum atoms in RbTaO$_3$. After substituting one or two sulfur atoms, the covalent bond between Ta and O remains intact, while a new covalent bond between Ta and S atoms forms.

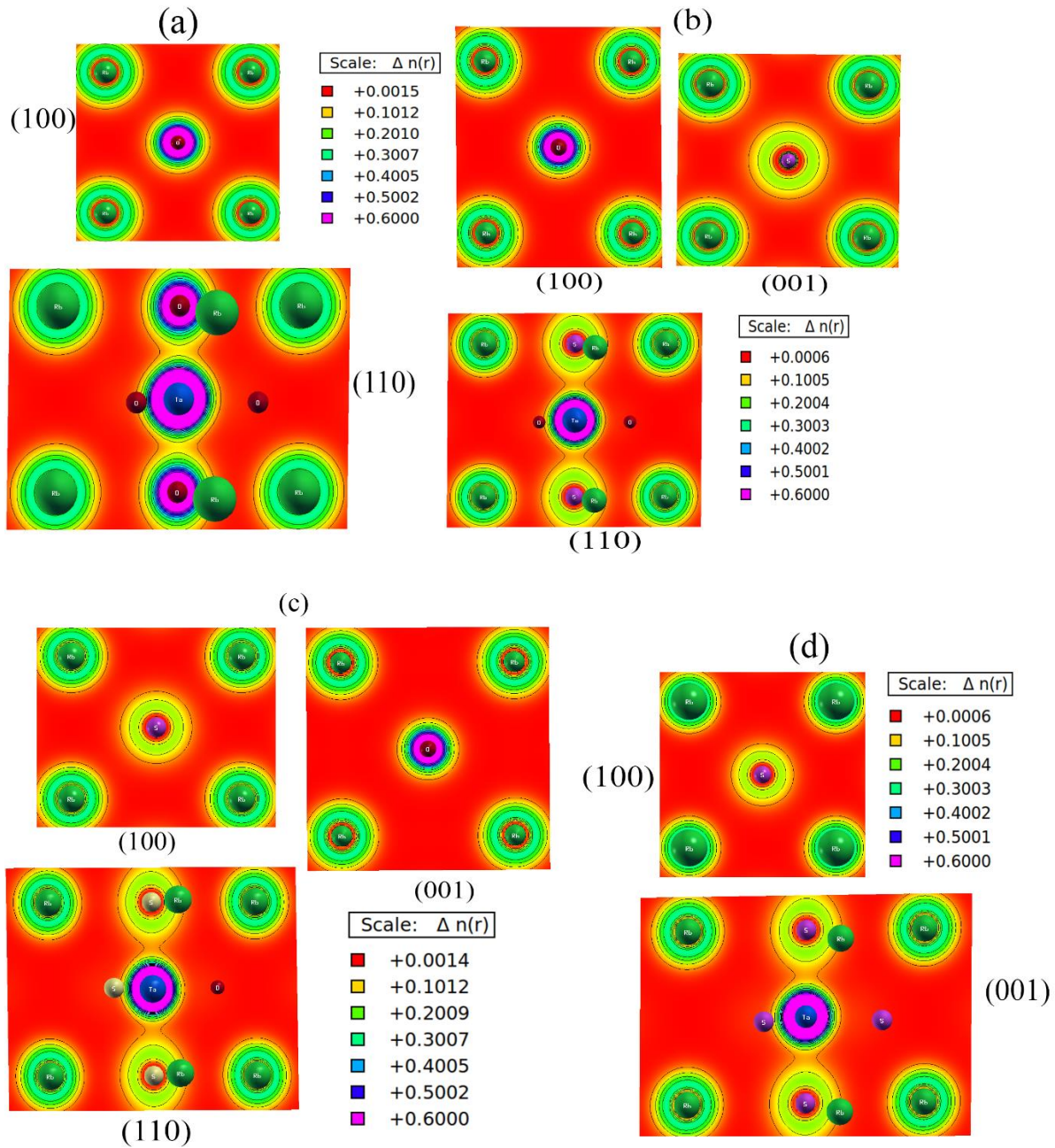

**Fig. 6:** Charge density plots for (a) $RbTaO_3$ (b) $RbTaO_2S$, (c) $RbTaOS_2$, and (d) $RbTaS_3$.

However, the covalent bond between Ta and O atoms is stronger than the newly formed Ta-S bond. When three sulfur atoms are substituted for oxygen, a very weak covalent bond between Ta and S is formed, significantly affects the compound's mechanical properties. Specifically, the compound exhibits lower hardness [supplementary section] and melting temperature [section 3.4] than the previous compounds with one or two sulfur substitutions. This is due to the reduced strength of the covalent bond between Ta and S atoms caused by three sulfur atoms. The weak bond also contributes to lower stability and weaker intermolecular interactions, which can affect the overall thermo-mechanical properties of the compound. Therefore, the choice of substitution atoms can significantly impact the properties of the resulting compound.

The Linus Pauling scale is used to assign electronegativity values to the constituent elements, with values of 3.5, 1.5, 0.8, and 2.58 for O, Ta, Rb, and S, respectively [63]. The difference in electronegativity between the atoms that form a covalent bond significantly impacts the bond's strength. In this case, since O has the highest electronegativity value among the elements considered, it has a greater tendency to attract electrons towards itself and develop a partial negative charge.

Therefore, the electronegativity difference between Ta and O is more significant than that between Ta and S. This is because the difference in electronegativity between Ta and O is greater than that between Ta and S, resulting in a greater polarization of the Ta-O bond, and, the covalent bond between Ta and O is stronger than that between Ta and S. Hence, we can conclude that the Ta-O bond is stronger than the Ta-S bond due to the higher electronegativity of O compared to S.

### *3.3 Optical Properties*

This section will explore how the substitution of S affects the optical properties of perovskite semiconductors, specifically $RbTaO_{3-x}S_x$. The optical constants, for example, dielectric function, $\varepsilon$, absorption coefficient ($\alpha$), reflectivity ($R$), loss function ($L$), refractive index ($n$), and extinction coefficient ($k$), play a vital role in understanding the nature of semiconductors. Determining the above-mentioned constants is also essential to investigate their optical characteristics for various applications.

It is important to highlight that the optical properties of two tetragonal phases, RbTaO$_2$S and RbTaOS$_2$, are significantly affected due to the phase transition from cubic to tetragonal symmetry. Furthermore, these properties are different for the *xx* and *zz* directions due to the symmetry change. Therefore, the impact of S substitution on the optical properties of semiconductors must be carefully examined in both directions to understand their properties better.

The frequency-dependent optical properties of semiconductors were analyzed by calculating their dielectric functions. The dielectric function ($\varepsilon$) is a well-known parameter used to describe the optical behavior of solid materials. It consists of two components, namely the real part ($\varepsilon_1(\omega)$) and the imaginary part ($\varepsilon_2(\omega)$), which were both taken into account in the study [64] as follows: $\varepsilon(\omega) = \varepsilon_1(\omega) + i\varepsilon_2(\omega)$ (3)

The real part ($\varepsilon_1(\omega)$) of the dielectric function is associated with the polarization and dispersion behavior of electromagnetic radiation, while the imaginary part ($\varepsilon_2(\omega)$) corresponds to the absorptive behavior of the material. These two components are interconnected through the Kramers-Kronig relations [65]. The equation used to determine the real part ($\varepsilon_1(\omega)$) of the dielectric function from the imaginary part is as follows [66]:

$$\varepsilon 1(\omega) = 1 + \frac{2}{\pi} p \int_0^\infty d\omega' \frac{\omega' \varepsilon_2(\omega')}{\omega'^2 - \omega^2} \quad (4)$$

where the principal value of the integral is expressed by P, the dielectric function's imaginary component ($\varepsilon_2(\omega)$) explains the transitions between occupied and unoccupied states for a fixed Brillouin zone k-vector. This quantity depends on the momentum matrix element and the density of states. The following expression can be used to calculate $\varepsilon_2(\omega)$ [66, 67]:

$$\varepsilon 2(\omega) = \left(\frac{h^2 e^2}{\pi m^2 \omega^2}\right) \sum_{i\,j} \int d^3 k <i_k|p^\alpha|j_k><j_k|p^\beta|i_k> x^\delta (\varepsilon_{i_k} - \varepsilon_{j_k} - \omega) \quad (5)$$

The moment matrix element (p) represents the interaction between the band α and β states at the crystal momentum k. The crystal wave vector k represents the wave functions corresponding to the valence and conduction bands as i$_k$ and j$_k$, respectively.

By utilizing the dielectric function, we can determine several crucial optical parameters, including the absorption coefficient ($\alpha(\omega)$), refractive index ($n(\omega)$), and extinction coefficient ($k(\omega)$).

$$\alpha(\omega) = \sqrt{2}\omega \left(\sqrt{\varepsilon_1^2(\omega) + \varepsilon_2^2(\omega)} - \varepsilon_1(\omega)\right)^{1/2} \quad (6)$$

$$n(\omega) = \frac{1}{\sqrt{2}} \left(\sqrt{\varepsilon_1^2(\omega) + \varepsilon_2^2(\omega)} + \varepsilon_1(\omega)\right)^{1/2} \quad (7)$$

$$k(\omega) = \frac{1}{\sqrt{2}}\left(\sqrt{\varepsilon_1^2(\omega) + \varepsilon_2^2(\omega)} - \varepsilon_1(\omega)\right)^{1/2} \tag{8}$$

The reflectance and optical conductivity can be deduced as given below

$$R(\omega) = \left|\frac{\sqrt{\varepsilon(\omega)}-1}{\sqrt{\varepsilon(\omega)}+1}\right|^2 \tag{9}$$

$$\sigma(\omega) = -\frac{i\omega}{4\pi}\varepsilon(\omega) \tag{10}$$

The calculated real ($\varepsilon_1(\omega)$) and imaginary ($\varepsilon_2(\omega)$) parts of the dielectric function are presented in Fig. 7 in the energy a range of 0-25 eV. The static real dielectric constants ($\varepsilon_1(0)$) are 4.2, 4.6, 6.79, 9.2, 5.2, and 11.4 for $RbTaO_3$, $RbTaO_2S$ [100], $RbTaO_2S$ [001], $RbTaOS_2$ [100], $RbTaOS_2$ [001], and $RbTaS_3$, respectively. One can observe that $RbTaS_3$ exhibits the highest value while $RbTaO_3$ exhibits the lowest value at the static point [$RbTaS_3$ > $RbTaOS_2$ > $RbTaO_2S$ > $RbTaO_3$].

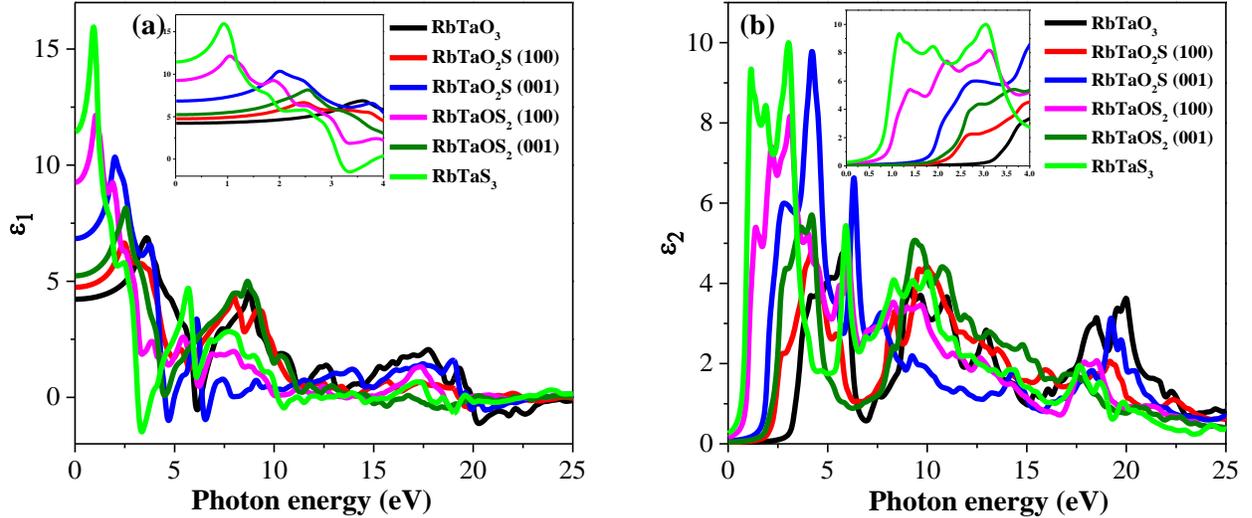

**Fig. 7:** The calculated (a) real dielectric constant, (b) imaginary dielectric constant of $RbTaO_{3-x}S_x$ compounds.

As depicted in Fig.7 (a), $\varepsilon_1(\omega)$ reaches its maximum value of 6.9 at 3.56 eV, 6.75 at 2.5 eV, 10.2 at 1.2 eV, 12 at 1.17 eV, 16 at 0.8 eV and 16.03 at 0.91eV for $RbTaO_3$, $RbTaO_2S$ [100], $RbTaO_2S$ [001], $RbTaOS_2$ [100], $RbTaOS_2$ [001], and $RbTaS_3$ respectively. Generally, negative real dielectric tensors indicate inadequate light transmission through the material and high reflection [68]. All these compounds exhibit negative values again after the 20 eV energy range, in this particular energy range and the materials behave like metals [69,70].

The imaginary part of the dielectric function tensor, $\varepsilon_2(\omega)$, is shown in Fig. 7 (b). It is notable from the figures that the static $\varepsilon_2(0)$ increases with the band gap energy; this is closely associated with the band structure. Hence, it remains zero until the band gap energies of these phases are reached. After this point, $\varepsilon_2(\omega)$ increases abruptly and reaches maximum values of 4.85 at 5.85 eV, 4.76 at 4.21 eV, 9.82 at 4.24 eV, 8.20 at 3.13 eV, 5.71 at 4.19 eV, and 10 at 3 eV for $RbTaO_3$, $RbTaO_2S$ [100], $RbTaO_2S$ [001], $RbTaOS_2$ [100], $RbTaOS_2$ [001], and $RbTaS_3$ compounds, respectively. It is worth noting that the maximum interaction range is almost the same as the plasmonic frequency range (4-6eV). Fig. 7(b) shows that, in all materials, the absorption of incident light falling on the surface of the material can be inferred from the peak values of $\varepsilon_2(\omega)$.

The absorption coefficient $\alpha(\omega)$ is closely related to the dielectric function [71]. The plasmon frequency is related to the energy loss function. The correlation between frequency and incident light is responsible for plasma resonance. The investigated materials, as well as other semiconductor materials, exhibit a value of zero for $\alpha(\omega)$ within the energy band gap.

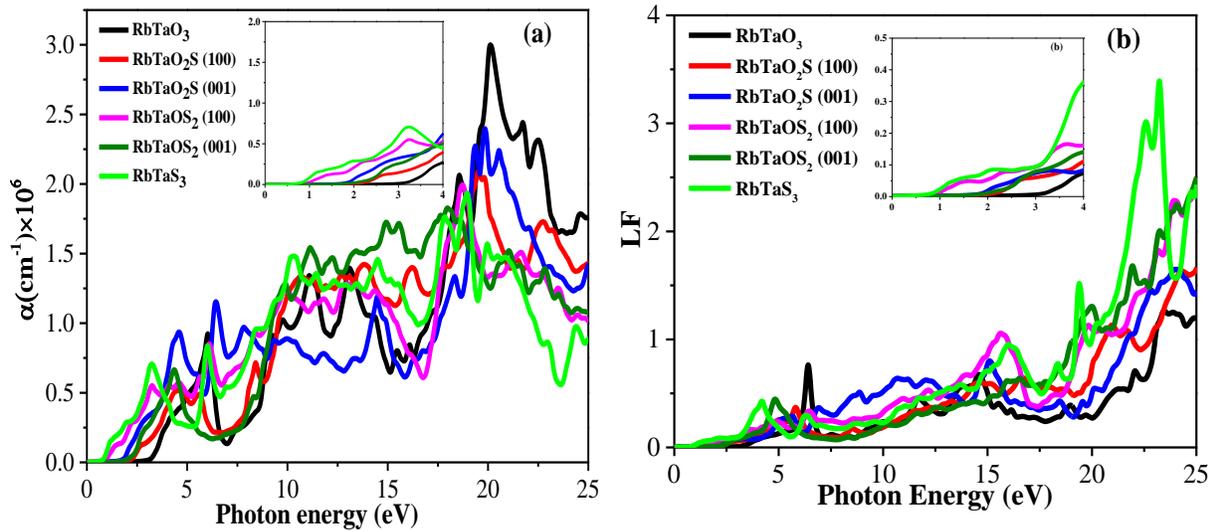

**Fig. 8:** The calculated (a) absorption coefficient and (b) loss function of $RbTaO_{3-x}S_x$ compounds.

The absorption coefficient $\alpha(\omega)$ in all of the studied compounds is found to be $10^6$ cm$^{-1}$ at their respective absorption edges, which occur at 2.7 eV, 1.6 eV, 1.4 eV, 1.7 eV, 0.7 eV, and 0.4 eV for $RbTaO_3$, $RbTaO_2S$ [100], $RbTaO_2S$ [001], $RbTaOS_2$ [100], $RbTaOS_2$ [001], and $RbTaS_3$, respectively, as shown in Fig. 8(a). The highest absorption values are observed in the 16 to 22 eV energy range, with $RbTaO_3$ exhibiting the highest absorption range among the studied compounds (RbTaO$_3$> RbTaO$_2$S [001] > RbTaO$_2$S [100] > RbTaOS$_2$ [100] > RbTaS$_3$> RbTaOS$_2$ [001]). Therefore, all of these cubic and tetragonal phases are found to be very efficient absorbers in the visible and ultraviolet range. The absorption coefficient of RbTaO$_2$S exhibits a remarkable peak of absorption, comparable to well-known high-efficiency solar cell perovskite materials such as CsPbI$_3$, CH$_3$NH$_3$PbCl$_3$, and MAPbCl$_3$. This makes RbTaO$_2$S a highly promising candidate for use in high-performance solar cell applications [72-74].

The energy loss function $L(\omega)$ characterizes the dissipation of energy within the compound resulting from electron-electron interactions and photoelectron interactions, which can be attributed to plasma losses [75]. Fig. 8 (b) shows the energy loss functions for RbTaO$_{3-x}$S$_x$ compounds. The energy loss begins at around 1 to 3.5 eV for all calculated compounds. With an increase in photon energy, the optical energy loss gradually increases and reaches its highest peak at around 20 eV for RbTaS$_3$. Among all the calculated compounds, RbTaO$_3$ exhibits the lowest electron loss in the higher energy range compared to the other compounds.

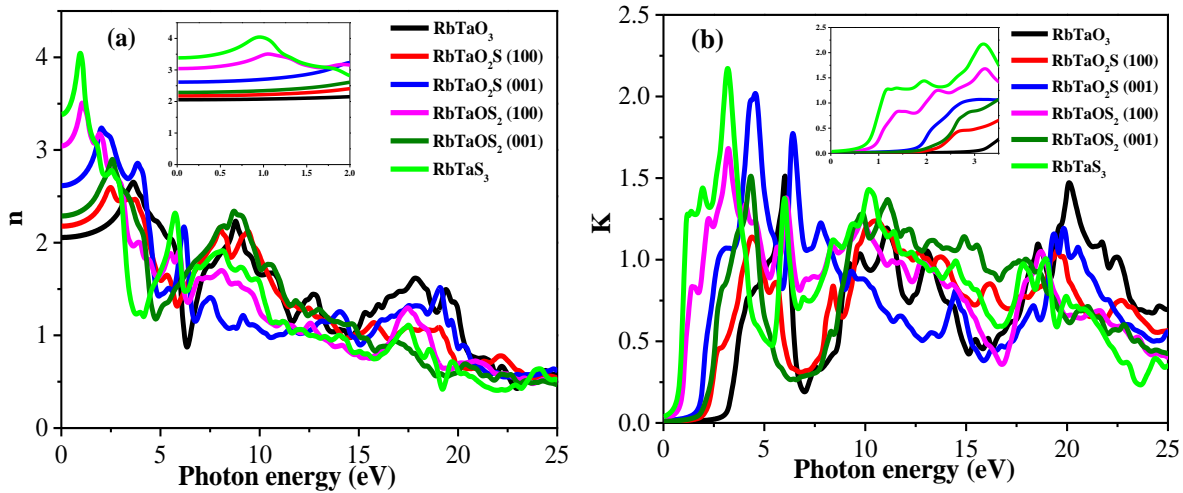

**Fig. 9:** Calculated (a) refractive index and (b) extinction coefficient of RbTaO$_{3-x}$S$_x$.

The extinction coefficient $k(\omega)$ and refractive index $n(\omega)$ of a material describe its ability to absorb incident photons and its velocity, respectively. The $n(\omega)$ of RbTaO$_{3-x}$S$_x$ is depicted in Fig. 9(a), and it is noticeable that its trend at different energy ranges is almost similar to the real dielectric tensor. The calculated values for the static point $n(0)$ are: 2.07, 2.16, 2.37, 3.05, 2.31, and 3.38 for RbTaO$_3$, RbTaO$_2$S [100], RbTaO$_2$S [001], RbTaOS$_2$ [100], RbTaOS$_2$ [001], and RbTaS$_3$, respectively. It is conspicuous that the $n(\omega)$ gradually increases with rising energy up to 2.6 at 3.6 eV, 2.6 at 2.47 eV, 3.2 at 1.9 eV, 3.49 at 0.9 eV, 2.9 at 2.7 eV, and 4.05 at 0.9 eV for

$RbTaO_3$, $RbTaO_2S$ [100], $RbTaO_2S$ [001], $RbTaOS_2$ [100], $RbTaOS_2$ [001], and $RbTaS_3$, respectively, and then decrease again at higher energy ranges. In general, semiconductor materials with wide band gaps possess lower value of refractive index at static point, which is exactly in agreement with our calculated result, where $RbTaO_3$ has the highest band gap and, at the same time, lowest refractive index. Similarly, the band gap of $RbTaS_3$ is the lowest, and its $n(\omega)$ is the highest among the studied compositions.

Fig. 9(b) shows the plot of the extinction coefficient. The $k(\omega)$ values for $RbTaO_{3-x}S_x$ remain at 0 in the energy band gap range. All compounds' maximum values are reached at around 3 eV to 6.5 eV when the light frequency becomes greater than the energy band gap. However, in the [100] directions of $RbTaO_2S$, the maximum peak is observed at 10.5 eV. It is seen from the figure that the spectra of $k(\omega)$ closely follow the imaginary dielectric tensor $\varepsilon_2(\omega)$.

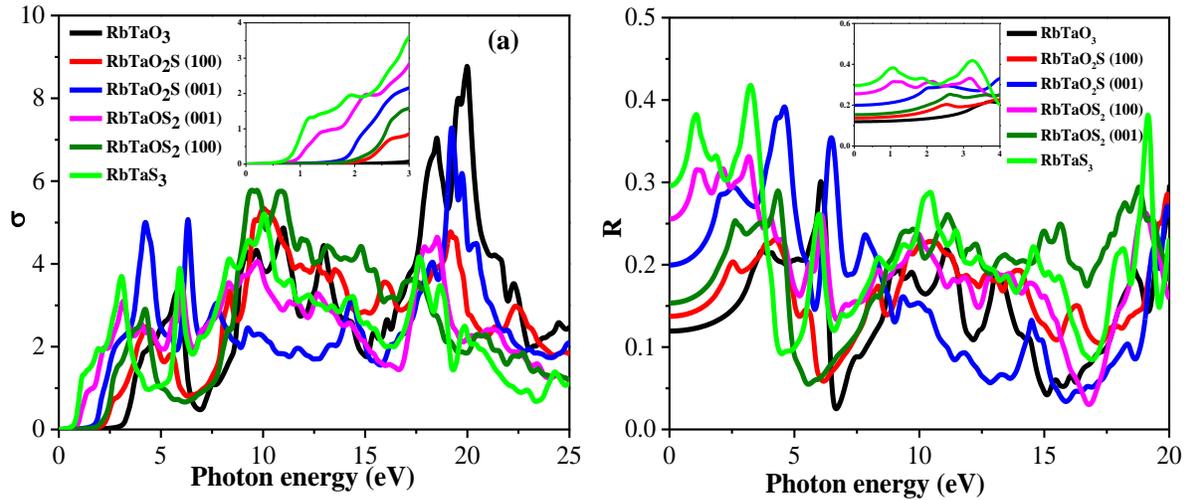

**Fig. 10:** Calculated (a) optical conductivity and (b) reflectivity of $RbTaO_{3-x}S_x$.

The optical conductivity of a solid is a measure of the degree of electron transport affected by light and is directly related to the refractive index and absorption coefficient. All cases studied here, display that the optical conductivity is zero at the energy band gap, as it should be. Upon reaching the threshold frequency, the optical conductivity increases up to the values of 3.37, 2.48, 5.02, 3.12, 2.98, and 3.73 for $RbTaO_3$, $RbTaO_2S$ [100], $RbTaO_2S$ [001], $RbTaOS_2$ [100], $RbTaOS_2$ [001], and $RbTaS_3$, respectively, as the first peak. Afterward, the optical conductivity decreases abruptly around the energy range of 4-7 eV for all the compounds, which is associated with the plasmon frequency. The optical conductivity peaks at 8.80 eV, 5.36 eV, 7.31 eV, 4.69 eV, 5.8 eV, and 5.29 eV for the RbTaO$_3$, RbTaO$_2$S [100], RbTaO$_2$S [001], RbTaOS$_2$ [100], RbTaOS$_2$ [001], and RbTaS$_3$ compounds, respectively.

The static values of reflectivity for these compounds are 0.12, 0.13, 0.20, 0.25, 0.15, and 0.29 for RbTaO$_3$, RbTaO$_2$S [100], RbTaO$_2$S [001], RbTaOS$_2$ [100], RbTaOS$_2$ [001], and RbTaS$_3$, respectively. The reflectivity's are poor – implying that these materials have potential to be used as anti-reflection coatings. The study found that the optical properties of RbTaO$_2$S are comparable to those of CsPbI$_3$ [76] along with their band gap values, which makes it a promising material for optoelectronic applications, such as solar cells. The results suggest that the calculated compounds have the potential to be utilized in solar cell technology.

### *3.4 Thermal Properties*

This section examines several critical thermodynamic parameters of RbTaO$_{3-x}$S$_x$, such as the Debye temperature, minimum thermal conductivity, lattice thermal conductivity, Grüneisen parameter, and melting temperature.

The Debye temperature $\Theta_D$ is a parameter that represents the temperature of the highest normal mode of vibration in a crystal. The Debye temperature can be estimated using the average sound velocity ($V_m$), which depends on the elastic constant (the bulk modulus $B$ and the shear modulus $G$). Debye temperature can be calculated by using the expression [77]:

$$\Theta_D = \frac{h}{k_B}[(\frac{3n}{4\pi})N_A\rho/M]V_m \qquad (11)$$

In this equation, the parameters *h, k$_B$, n, N$_A$, ρ*, and *M* correspond to Planck's constant, Boltzmann's constant, the number of atoms in the molecule, Avogadro's number, the mass density, and the molar mass, respectively. The following equation can estimate the average sound velocity in polycrystalline material:

$$Vm = [\frac{1}{3}(\frac{1}{v_l^3} + \frac{2}{v_t^3})]^{-1/3} \qquad (12)$$

Where, the longitudinal sound velocity ($v_l$) and transverse sound velocity ($v_t$) in a solid are related to the elastic moduli (bulk and shear modulus) and density. Using Navier's equation we get:

$$V_l = [(3B + 4G)/3\rho]^{1/2} \text{ and } V_t = [G/\rho]^{1/2} \qquad (13)$$

The longitudinal sound velocity has much higher value than the transverse sound velocity [78], it requires more energy to vibrate neighboring atoms during the propagation of a transverse wave. This decrease in wave energy leads to a reduced wave velocity. The calculated values of Debye temperature ($\Theta_D$), displayed in Fig. 11, show variations in line with mechanical properties. Our results indicate that S-substitution leads to a decrease in the Debye temperature.

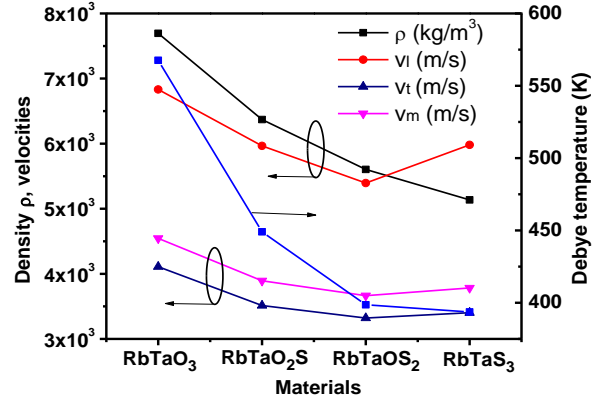

**Fig.11:** Debye temperatures of the RbTaO$_{3-x}$S$_x$ compounds.

Generally, harder solids have higher $\Theta_D$, whereas a low value of $\Theta_D$ indicates a low value of minimum thermal conductivity as well as lattice thermal conductivity [79]. The highest $\Theta_D$ value is found for RbTaO$_3$, while the lowest is for RbTaS$_3$. The order of $\Theta_D$ values can be written as follows: RbTaO$_3$> RbTaO$_2$S > RbTaOS$_2$> RbTaS$_3$. The average sound velocity of these compounds also follows the same trend.

**Lattice thermal conductivity**

The empirical formula for determining the lattice thermal conductivity ($K_{ph}$) of RbTaO$_{3-x}$S$_x$, based on the heat conduction resulting from the vibration of the lattice ions in a solid, was derived by Slack [80] and can be expressed as:

$$K_{ph} = A(\gamma) \frac{M_{av}\Theta_D^3}{\gamma^2 n^{2/3} T} \qquad (14)$$

$$\gamma = \frac{3(1+\upsilon)}{2(2-3\upsilon)} \qquad (15)$$

where the anharmonicity of phonon is represented by the Grüneisen parameter, $\gamma$. Materials that exhibit high values of the Grüneisen parameter ($\gamma$) typically have a substantial amount of anharmonic contributions, resulting in low phonon thermal conductivity. The results of this study have indicated that all compositions of RbTaO$_{3-x}$S$_x$ possess a moderate value of the Grüneisen parameter ($\gamma$), as shown in Table 3. After calculating the Grüneisen parameter, the coefficient $A(\gamma)$ can be estimated using the following formula:

$$A(\gamma) = \frac{4.85628 \times 10^7}{2(1 - \frac{0.514}{\gamma} + \frac{0.228}{\gamma^2})} \tag{16}$$

The modified Clarke's model [81] provides an expression for the theoretical minimum intrinsic thermal conductivity, given as:

$$K_{min} = K_B V_m \left(\frac{M}{n\rho N_A}\right)^{-\frac{2}{3}} \tag{17}$$

Table 3 presents the calculated values of the minimum thermal conductivity, $K_{min}$, lattice thermal conductivity, $K_{ph}$ at 300 K, and the Grüneisen parameter, $\gamma$, for RbTaO$_{3-x}$S$_x$. As seen in Table 3, the thermal conductivity of the S-substituted compositions is much lower than that of RbTaO$_3$, thermal conductivity is inversely related with ZT values of a TE material. Lowering the thermal conductivity is a strong indication of enhanced thermoelectric properties for S-substituted compositions, like AgSbTe$_2$ ($K_{ph}$ ~ 0.39 Wm$^{-1}$K$^{-1}$, $ZT$ ~ 1.6 at 673 K), Ag$_9$TlTe$_5$ ($K_{ph}$ ~ 0.22 Wm$^{-1}$K$^{-1}$, $ZT$ ~ 1.2 at 700 K), BiCuSeO ($K_{ph}$ ~ 0.45 Wm$^{-1}$K$^{-1}$, $ZT$ ~ 0.9 at 923 K)[82-84]. Fig. 12 (a) shows the temperature dependence of $K_{ph}$ for RbTaO$_{3-x}$S$_x$. The results indicate that $K_{ph}$ exhibits a sharp decrease in the 0-1000 K, followed by a gradual decrease between 1000-2000 K, and ultimately attains a constant value at elevated temperatures.

The melting temperature of perovskite semiconductors is an important property that can impact their thermal stability and suitability for various applications, particularly those which involve high-temperature processing or operation. The theoretical estimation of the melting temperature ($T_m$) for RbTaO$_{3-x}$S$_x$ has been conducted using the following equations [85]:

For cubic phases:

For cubic phases:

$$T_m(K) = 553 + (5.911)C_{11} \tag{18}$$

For tetragonal phases:

$$T_m(K) = 3C_{11} + (1.5)C_{33} + 354 \tag{19}$$

where, $C_{11}$ and $C_{33}$ denote the elastic constants [presented in supplementary section]. The obtained values for $T_m$ are presented in Table 3. Comparing the $T_m$ values, it can be observed that the $T_m$ of RbTaO$_3$ is greater than those of RbTaO$_2$S, RbTaOS$_2$, and RbTaS$_3$ compounds. This observation is consistent with Young's modulus [supplementary section] since $T_m$ has a strong correlation with it. We calculated the melting temperatures of RbTaO$_3$, RbTaO$_2$S, RbTaOS$_2$, and RbTaS$_3$ under ambient conditions and obtained the values are 2953 K, 1471 K, 1201 K, and 1889 K, respectively.

**Specific heat and thermal expansion coefficient:**

Specific heat is a significant thermal property that plays a crucial role in casting and heat treatment processes. It dictates the amount of heat necessary for the operations. Heat capacity measures a substance's ability to retain heat. Supplying heat to a material always results in a corresponding material specific temperature increase.

For temperatures above 300 K, the quasi-harmonic Debye model provides an accurate explanation of the specific heat at constant volume ($C_V$). It takes into account the lattice vibrations of atoms and the anharmonicity of these vibrations, treating them as a collection of coupled harmonic oscillators [86-89] with variable volume.

$$C_V = 9nN_A K_B \left(\frac{T}{\theta_D}\right) \int_0^{\theta_D} dx \frac{x^4}{(e^x-1)^2} \qquad (20)$$

where, N$_A$, n, $k_B$ be Avogadro's number, the number of atoms per formula unit, and Boltzmann constant, respectively. x$_D$ is defined as $\theta_D/T$. The linear thermal expansion coefficient ($\alpha$) and specific heat at constant pressure ($C_P$) are calculated using the equations given below [86]:

$$\alpha = \frac{\gamma C_v}{3 B_T v_m}, \text{ and } C_p = C_v(1 + \alpha \gamma T) \qquad (21)$$

where, $B_T$ and $v_m$ be the isothermal bulk modulus and molar volume.

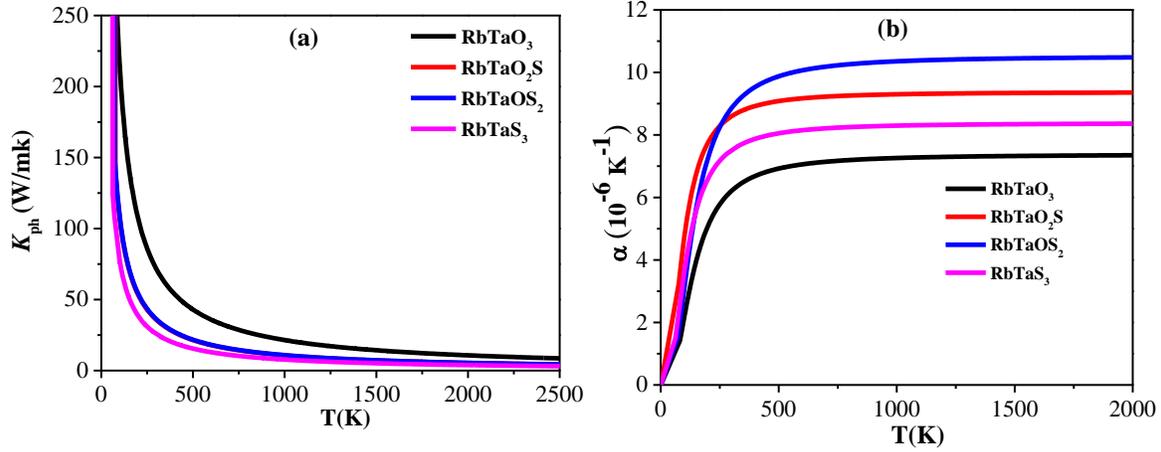

**Fig. 12:** (a) Temperature-dependent lattice thermal conductivity, K$_{ph}$ (W/mK), and (b) linear thermal expansion coefficient (α) of RbTaO$_{3-x}$S$_x$.

Fig. 12 (b) illustrates the relationship between temperature and the coefficient of thermal expansion (α) for RbTaO$_{3-x}$S$_x$ solid solutions. Our research shows that α rapidly increases until it reaches a specific temperature for all compounds around 600 K to 800 K. Afterward, α gradually increases, reaching an almost constant value. At a temperature of 300 K, the coefficients of thermal expansion for these solid solutions are $6.18\times10^{-5}$ K$^{-1}$ for RbTaO$_3$, $8.58\times10^{-5}$ K$^{-1}$ for RbTaO$_2$S, $8.82\times10^{-5}$ K$^{-1}$ for RbTaOS$_2$, and $7.40\times10^{-5}$ K$^{-1}$ for RbTaS$_3$.

**Table 3:** Calculated melting temperature ($T_m$), Grüneisen parameter (γ), lattice thermal conductivity ($K_{ph}$), minimum thermal conductivity ($K_{min}$), specific heat ($C_v$, $C_p$) and the thermal expansion coefficient (α) of RbTaO$_{3-x}$S$_x$ perovskite compounds.

| Phases | $T_m$ | γ | $K_{ph}$ (W/mK) | $K_{min}$ (W/mK) | $C_V$ | $C_P$ | α ($10^{-6}$/K) |
|---|---|---|---|---|---|---|---|
| RbTaO$_3$ | 2952.95 | 1.349 | 72.31 | 1.10 | 104.55 | 104.81 | 6.18 |
| RbTaO$_2$S | 1470.88 | 1.426 | 36.11 | 0.80 | 114.10 | 114.52 | 8.58 |
| RbTaOS$_2$ | 1201.60 | 1.263 | 36.56 | 0.67 | 104.55 | 104.89 | 8.82 |
| RbTaS$_3$ | 1889.01 | 1.549 | 27.54 | 0.63 | 110.07 | 110.42 | 7.40 |

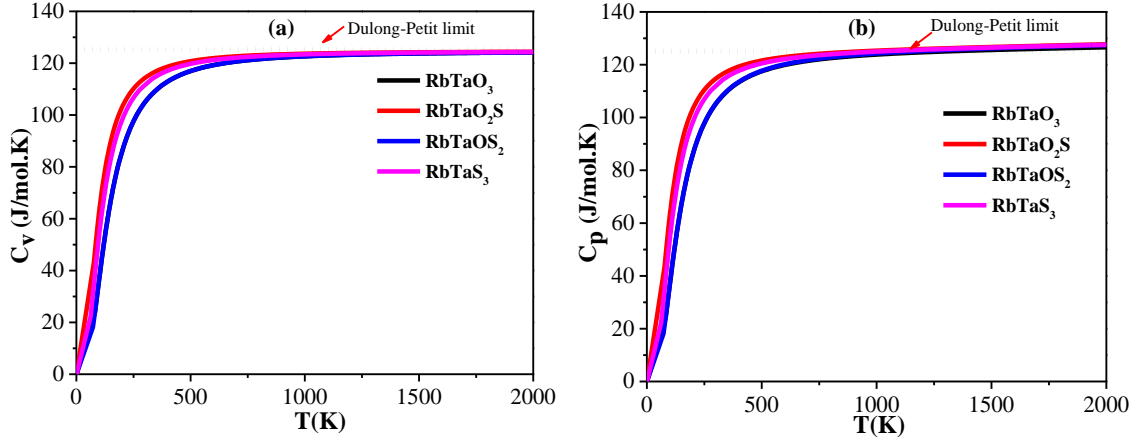

**Fig. 13:** The temperature-dependent specific heats for the RbTaO$_{3-x}$S$_x$ solid solutions are reported in two forms: (a) $C_v$ at constant volume and (b) $C_p$ at constant pressure.

In Fig. 13, we can observe the temperature dependence of specific heats, $C_v$ and $C_p$, within the range of 0-2000 K. The reason for the increase in specific heats with the increase in temperature is thermal softening. At temperatures lower than 600 K, the heat capacities follow the Debye T$^3$ power-law, whereas at temperatures exceeding 1000 K, they nearly conform to the Dulong-Petit (DP) model. It is common for solid materials that although $C_v$ approaches the DP limit [90] at higher temperatures (1000 K), the DP model slightly underestimates $C_p$ at these temperatures. However, the deviations of $C_p$ values from the DP model for all titled compounds at 2000 K were always less than around 2.9-3.9%. Table 3 presented our calculated values of $C_v$ and $C_p$ for RbTaO$_{3-x}$S$_x$. Our calculations indicate that the compounds we analyzed, especially RbTaO$_3$, exhibit high values of lattice thermal conductivity, suggesting their potential applications as heat sink materials. This is similar to another material, Ti$_2$BC [91], which have been predicted to have heat sink properties.

## 4. Conclusion

This paper systematically investigates the stability, electronic, optical, mechanical, and thermal properties of RbTaO$_{3-x}$S$_x$ Perovskite semiconducting compounds using first-principles methods. The substitution of S for O leads to a structural transition from cubic to tetragonal for RbTaO$_2$S, and RbTaOS$_2$, which then retain back to cubic phase for the composition RbTaS$_3$. The formation energy and phonon dispersion curves revealed their phase stability. Electronic band structure and DOS results confirmed the semiconducting nature, with a clear band gap between the valence and conduction bands around the Fermi level. The reduction of band gap is found due to S substitution at the O sites. Presence of ionic and covalent bonds is confirmed from the partial DOS and charge density plots. The optical properties results agree well with the band structure results [confirming semiconducting nature with different band gaps]. The high absorption coefficient (in excess of $10^6$ cm$^{-1}$) makes all the compounds very good absorbing materials. Furthermore, very low reflectivity of these compositions also supports the high absorption possibility for them. Particularly, RbTaO$_2$S is expected to have high light absorption [92-94], making it an attractive material for use in the solar cell technology. Anisotropic optical properties are found for RbTaO$_2$S and RbTaOS$_2$ phases; both of which are tetragonal. The elastic constants and moduli are also decreased due to S substitution. The elastic properties reveal that RbTaO$_3$, RbTaO$_2$S, and RbTaOS$_2$ are brittle, while RbTaS$_3$ exhibits ductile behavior. We plotted anisotropy graphics of the mechanical properties in both 3D and 2D [supplementary section]. Based on these graphics, the elasticity of RbTaO$_{3-x}$S$_x$ compounds are found to be considerably anisotropic. Like elastic properties, the Debye temperature, thermal conductivity, melting temperature, specific heat capacity, and thermal expansion coefficient are also significantly lowered by S substitution. Lowering of thermal conductivity along with the band gap values makes S-containing phases as promising thermoelectric materials. The results of our study provide valuable insights and should serve as a useful reference for future experimental research.

**Declaration of interests**

The authors declare that they have no known competing financial interests or personal relationships that could have appeared to influence the work reported in this paper.

**Acknowledgments**

This work was carried out with the aid of a grant (grant number: 21-378 RG/PHYS/AS_G - FR3240319526) from UNESCO-TWAS and the Swedish International Development Cooperation Agency (SIDA). The views expressed herein do not necessarily represent those of UNESCO-TWAS, SIDA or its Board of Governors.

**CRediT Author contributions**

**(Supplementary)**

# Oxysulfide Perovskites: Reduction of the Electronic Band Gap of RbTaO$_3$ by Sulfur Substitution to Enhance Prospective Solar Cell and Thermoelectric Performances

H. Akter[1,2], M. A. Ali[1,2,*], M. M. Hossain[1,2], M. M. Uddin[1,2], S. H. Naqib[2,3]

[1]Advanced Computational Materials Research Laboratory, Department of Physics, Chittagong University of Engineering and Technology (CUET), Chattogram-4349, Bangladesh

[2]Department of Physics, Chittagong University of Engineering and Technology (CUET), Chattogram-4349, Bangladesh

[3]Department of Physics, University of Rajshahi, Rajshahi-6205, Bangladesh

## 1. Mechanical Properties

The elastic properties of solids describe the ability of a material to deform when subjected to stresses of different types and to return to its original shape when the stresses are removed. Study of elastic properties yields dependable results regarding the mechanical properties of a material, including its stability, strength, ductility/brittleness, Young's modulus ($Y$), bulk modulus ($B$), shear modulus ($G$), anisotropy factor, and other elastic parameters [1]. We started our inquiry of mechanical characteristics using the strain-stress approach to compute the independent elastic constants. Elastic constants and the Hill's approximation were utilized to determine the average values of the elastic moduli's upper and lower limits for polycrystalline materials [2]. Both elastic moduli and elastic constants are then used to estimate fracture toughness, ductility or brittleness, hardness, and anisotropic characteristics of the discussed perovskite systems. The obtained elastic constants of RbTaO$_{3-x}$S$_x$ are shown in Table S1. The mechanical stability criteria of cubic and tetragonal structures are presented as follows.

There are three basic accepted elastic stability criteria for cubic crystals [3].

$C_{11}+2C_{12}>0$, $C_{44}>0$, $C_{11}-C_{12}>0$. (S1)

The obtained results for RbTaO$_3$ and RbTaS$_3$ satify these criteria, which approve the mechanical stability of these compounds.

The six independent elastic constants obtained for the tetragonal crystal, i.e., $C_{11}$, $C_{12}$, $C_{13}$, $C_{33}$, $C_{44}$, and $C_{66}$ confirmed elastic stability criteria. The spinodal, shear, and Born criteria are used in connection with the bulk, shear, and tetragonal shear moduli, respectively. The following expressions characterize them [4].

$C_{11} - C_{12} > 0$,

$C_{11} + C_{33} - 2C_{13} > 0$,

$C_{11}>0$, $C_{33}>0$, (S2)

$C_{44}>0$, $C_{66}>0$,

$2C_{11} + C_{33} + 2C_{12} + 4C_{13} > 0$,

$\frac{1}{3}(C_{11} + 2C_{13}) < B < \frac{1}{3}(C_{11} + 2C_{33})$.

For tetragonal crystals, the resistances to linear compression are quantified by $C_{11}$ and $C_{33}$ along the x and z directions, respectively [5]. In equation (2) and equation (3) we can see that the elastic constants are all positive and $C_{11}$ is larger than $C_{12}$, and $2C_{13}$ is smaller than $C_{11}+C_{33}$, thereby indicating that these RbTaO$_{3-x}$S$_x$ compounds meet the mechanical stability criterion and these compounds are mechanically stable. Table S1 displays the calculated values, demonstrating their mechanical stability. Notably, our results exhibit a significant concurrence with previously calculated values, thereby substantiating the accuracy of our calculations [6].

**Table S1:** Calculated stiffness constants (GPa) of RbTaO$_{3-x}$S$_x$ compounds with cubic and tetragonal structures.

| Compound | $C_{11}$ | $C_{12}$ | $C_{13}$ | $C_{44}$ | $C_{33}$ | $C_{66}$ |
|---|---|---|---|---|---|---|
| RbTaO$_3$ | 406.0836 | 75.8167 | - | 110.6489 | - | - |
| RbTaO$_2$S | 241.1560 | 91.4681 | 42.5394 | 56.0044 | 262.2789 | 110.9909 |
| RbTaOS$_2$ | 184.2718 | 19.3252 | 30.8164 | 62.4726 | 196.5253 | 36.1300 |
| RbTaS$_3$ | 226.0608 | 43.6594 | - | 44.3736 | - | - |

Using these elastic constants, the bulk modulus $B$, the share modulus $G$, Poisson's ratio ($v$) and Young modulus $Y$, are estimated. In solid, with the Voigt-Reuss Hill (VRH) approximation, it is possible to describe an isotropic material by bulk and shear moduli [7, 8]. The bulk modulus and the shear modulus in the Hill's average for cubic [7] and tetragonal [9] systems are given as,

$$B_V = \frac{1}{3}(C_{11} + 2C_{12}), \tag{S3}$$

$$G_V = \frac{1}{5}[(C_{11} - C_{12}) + 3C_{44}]. \tag{S4}$$

And,

$$B_V = \frac{1}{9}(2C_{11} + C_{12}) + C_{33} + 4C_{13}, \tag{S5}$$

$$G_V = \frac{1}{30}(M + 3C_{11} - 3C_{12} + 12C_{44} + 6C_{66}). \tag{S6}$$

The above quantities can be taken in the Reuss average [8,9],

$$B_R = \frac{1}{3}C_{11} + 2C_{12}) = B_V, \tag{S7}$$

$$G_R = [\frac{4}{5}(C_{11} - C_{12})^{-1} + \frac{3}{5}C_{44}^{-1}]^{-1}. \tag{S8}$$

and,

$$B_R = \frac{C^2}{M} \tag{S9}$$

$$G_R = 15/[\frac{18B_V}{C^2} + \frac{6}{C_{11}-C_{12}} + \frac{6}{C_{44}} + \frac{3}{C_{66}}]. \tag{S10}$$

Where, $M = C_{11}+C_{12}+2C_{33}-4C_{13}$ and $C^2 = (C_{11} + C_{12})C_{33} - 2C_{13}^2$.

Also in Hill empirical average [2] the Bulk modulus and share modulus are expressed by,

$$B = \frac{1}{2}(B_V + B_R), \tag{S11}$$

$$G = \frac{1}{2}(G_V + G_R), \tag{S12}$$

respectively. The young modulus ($Y$) and Poisson's ($v$) ratio for polycrystalline materials [10] are frequently calculated by using the following relationships:

$$Y = \frac{9BG}{3B+G} \text{ and } v = \frac{3B-Y}{6B}. \tag{S13}$$

Young's modulus ($Y$), bulk modulus ($B$), and shear modulus ($G$) measure the hardness and stiffness of the materials. Further, to predict the hardness of solids, $C_{44}$ is the best indicator among the elastic constants [11]. Here, we see that RbTaO$_3$'s shear moduli, which signifies the material's hardness, has larger bulk values. Additionally, the stiffness is increased by the higher Young's modulus value. In Table S2, from our calculated values of other compounds, it is clear that when we substitute S for O, we see that the elastic moduli of RbTaO$_2$S, RbTaOS$_2$, and RbTaS$_3$ are decreasing compared to RbTaO$_3$. So, the hardness estimated for these compounds is hoped to be in the following order: RbTaO$_3$ > RbTaO$_2$S > RbTaOS$_2$ > RbTaS$_3$.

Pugh's ratio *B/G* helps the categorization of the brittleness and ductility of solids. The ratio *B/G* > 1.75, indicates that the material is ductile in nature; otherwise, (*B/G* < 1.75) it reveals brittleness [12]. Table S2 shows that RbTaO$_3$, RbTaO$_2$S, and RbTaOS$_2$ compounds are brittle, but RbTaS$_3$ is ductile. Another significant elastic parameter is the Poisson ratio v, which helps to confirm brittleness and ductility. If the Poisson ratio *v* > 0.26, according to Frantsevich's rule, it will present ductility; lower than this value suggests brittleness [13]. The computed values of v are found to be 0.21, 0.23, 0.19, and 0.26 for the RbTaO$_3$, RbTaO$_2$S, RbTaOS$_2$, and RbTaO$_3$ compounds, respectively. So, these values further verify the brittle nature of RbTaO$_3$, RbTaO$_2$S, and RbTaOS$_2$ compounds and ductile nature of RbTaS$_3$, in complete agreement with Pugh's ratio. The ratio $B/C_{44}$ gives the machinability index, an important mechanical performance indicator [14]. A material with low shear resistance and a small $C_{44}$ value tends to have high machinability. In the case of RbTaS$_3$, it exhibits higher machinability compared to the other compounds.

To calculate the hardness of these materials, the hardness parameters H$_{miao}$ and H$_{Chen}$ were obtained using the following formulae [15-16]:

$$H_{miao} = \frac{(1-2v)Y}{6(1+v)} \text{ and, } H_{chen} = 2[\left(\frac{G}{B}\right)^2 G]^{0.585} - 3 \qquad \text{(S14)}$$

According to the Chen's formula [16], superhard materials have a hardness value greater than 40 GPa. The computed values of H$_{chen}$ for RbTaO$_3$, RbTaO$_2$S, RbTaOS$_2$, and RbTaS$_3$ are 19.8, 12.39, 13.35, and 8.29 GPa, respectively. Thus, these materials are not superhard by any means.

**Table S2:** Computed bulk modulus $B$ (GPa), Young's modulus $Y$ (GPa), shear modulus $G$ (GPa), Pugh's ratio ($B/G$), Poisson's ratio $v$, machinability index $B/C_{44}$, and hardness parameters ($H_{miao}$ and $H_{chen}$) (GPa) for RTaO$_{3-x}$S$_x$ compounds.

| Compound | B | G | Y | B/G | v | B/C$_{44}$ | H$_{miao}$ | H$_{chen}$ |
|---|---|---|---|---|---|---|---|---|
| RbTaO$_3$ | 185.905 | 129.957 | 316.192 | 1.4305 | 0.216 | 1.68 | 24.61 | 19.68 |
| RbTaO$_2$S | 121.839 | 78.640 | 194.149 | 1.5493 | 0.234 | 2.17 | 13.95 | 12.39 |
| RbTaOS$_2$ | 80.643 | 61.818 | 147.710 | 1.3045 | 0.194 | 1.29 | 12.61 | 13.35 |
| RbTaS$_3$ | 104.459 | 59.473 | 149.959 | 1.7648 | 0.260 | 2.35 | 9.52 | 8.29 |

Elastic anisotropy is highly significant in comprehending the mechanical characteristics of a crystal under mechanical stress. Many low-symmetry crystals possess a high level of elastic anisotropy. In order to determine the elastic properties of these compounds in different crystallographic directions, their elastic anisotropic behaviors were investigated. The elastic anisotropy of crystals can be assessed using various formalisms. In the case of cubic crystals, the Zener anisotropy is used to measure their anisotropic behavior [17].

$$A^U = \frac{2C_{44}}{(C_{11}-C_{12})} \qquad \text{(S15)}$$

The value of 'A$^U$' being equal to unity implies the isotropic nature while 'A$^U$' less/greater than 1 shows an anisotropic character [17]. The obtained values of RbTaO$_3$ (0.67) and RbTaS$_3$ (0.48) are smaller than '1', indicating the anisotropic nature.

For the tetragonal crystal, shear anisotropy factors A$_{\{100\}}$, and A$_{\{001\}}$ for the {100} and {001} planes, respectively, were calculated by the following equation:

$$A_{100} = \frac{4C_{44}}{(C_{11}+C_{33}-2C_{13})} \text{ and } A_{001} = \frac{4C_{66}}{(C_{11}+C_{22}-2C_{12})} \quad \text{(S16)}$$

The factors $A_{100}$ and $A_{001}$, smaller or greater than unity is a measure of the degree of anisotropy, while the values being equal to unity reveal completely isotropic properties. Table S3 shows that the obtained values for tetragonal compounds RbTaO$_2$S and RbTaOS$_2$ exhibit an anisotropic nature.

By utilizing the relationship between the highest anisotropy (Voigt, $V$) and the lowest anisotropy (Reuss, $R$) of the bulk and shear modulus, it is possible to calculate the percentage anisotropy in compressibility ($A_B$), percentage shear anisotropy ($A_G$), and the universal anisotropy index $A^U$ [18],

$$A^U = 5\frac{G_V}{G_R} + \frac{B_V}{B_R} - 6, A_B \frac{B_V-B_R}{B_V+B_R} \times 100\% \times 100\%, \text{ and } A_G = \frac{G_V-G_R}{G_V+G_R} \times 100\% \quad \text{(S17)}$$

A non-zero value for A$_B$ and A$_G$ implies anisotropy; their zero values expose the isotropy of the crystal. The crystal is isotopic, when the universal anisotropic index (A$^U$) equals zero. The deviation from zero ascertains the level of elastic anisotropy. Therefore, our obtained universal anisotropy index ($A^U$) of 0.51 for RbTaO$_2$S$_1$ and 0.45 for RbTaO$_1$S$_2$ declare that the tetragonal phase of the materials is anisotropic.

**Table S3:** Calculated elastic anisotropy indices ($A^U$, $A_B$, A$_G$, $A_1$ and $A_3$) of RbTaO$_{3-x}$S$_x$.

| Phase | A$_1$ | A$_3$ | A$_B$ | A$_G$ | A$^U$ |
|---|---|---|---|---|---|
| RbTaO$_3$ | - | - | - | - | 0.67 |
| RbTaO$_2$S | 0.53 | 1.48 | 0.10 | 4.86 | 0.51 |
| RbTaOS$_2$ | 0.78 | 0.43 | 0.16 | 4.31 | 0.45 |
| RbTaS$_3$ | - | - | - | - | 0.48 |

To further explore the anisotropic characteristics exhibited by all the compositions being investigated, we have employed both two-dimensional (2D) and three-dimensional (3D) plots. These plots were derived by utilizing the ELATE code [19]. Furthermore, we included a summary of all the elastic properties, such as Young's modulus ($Y$), linear compressibility ($K=1/B$), shear modulus ($G$), and Poisson's ratio ($\rho$), along with their respective minimum and maximum values in Table S4. This table provides an overview of the elastic characteristics of the materials analyzed and allows for quick comparisons to be made between the different compositions. Here, only the 2D and 3D diagrams of the $RbTaO_3$ and $RbTaO_2S$ compounds are displayed in Fig. S1 (a-b) and Fig. S2. If the plot takes on a spherical (3D) or circular (2D) shape, it indicates a perfectly isotropic behavior in the 3D/2D. On the contrary, any departure from a spherical shape indicates the level of anisotropy in various directions within the three-dimensional space. The same applies for the 2D case where the anisotropy is within a particular crystal plane. The Young's modulus ($Y$), shear modulus ($G$), and Poisson's ratio ($v$) of materials exhibit significant anisotropy in all the planes. However, in the case of cubic $RbTaO_3$ and $RbTaS_3$, the linear compressibility of the material is isotropic and same in all directions, as depicted in Fig. S1(b). On the other hand, for $RbTaO_2S$ and $RbTaOS_2$, as shown in Fig. S2(b), the linear compressibility displays strong anisotropic behavior in the $xz$- and $yz$-planes, while it exhibits isotropic behavior in the $xy$-plane.

**Table S4:** The lower and upper limits of the Young's modulus, compressibility, shear modulus, and Poisson's ratio of $RbTaO_{3-x}S_X$ perovskites.

| Phases | $Y_{min.}$ (GPa) | $Y_{max.}$ (GPa) | $A_Y$ | $K_{min}$ $(TPa^{-1})$ | $K_{max}$ $(TPa^{-1})$ | $A_K$ | $G_{min.}$ (GPa) | $G_{max.}$ (GPa) | $A_G$ | $v_{min.}$ | $v_{max.}$ | $A_v$ |
|---|---|---|---|---|---|---|---|---|---|---|---|---|
| $RbTaO_3$ | 276.99 | 382.23 | 1.38 | 1.79 | 1.79 | 1.00 | 110.65 | 165.13 | 1.49 | 0.122 | 0.344 | 2.81 |
| $RbTaO_2S$ | 154.84 | 261.73 | 1.69 | 2.62 | 2.96 | 1.12 | 56.004 | 113.43 | 2.02 | 0.095 | 0.449 | 4.72 |
| $RbTaOS_2$ | 105.29 | 187.2 | 1.77 | 3.72 | 4.34 | 1.16 | 36.13 | 82.473 | 2.28 | 0.080 | 0.457 | 5.65 |
| $RbTaS_3$ | 116.61 | 211.93 | 1.81 | 3.191 | 3.191 | 1.00 | 44.374 | 91.201 | 2.05 | 0.100 | 0.480 | 4.78 |

(a) Young's Modulus

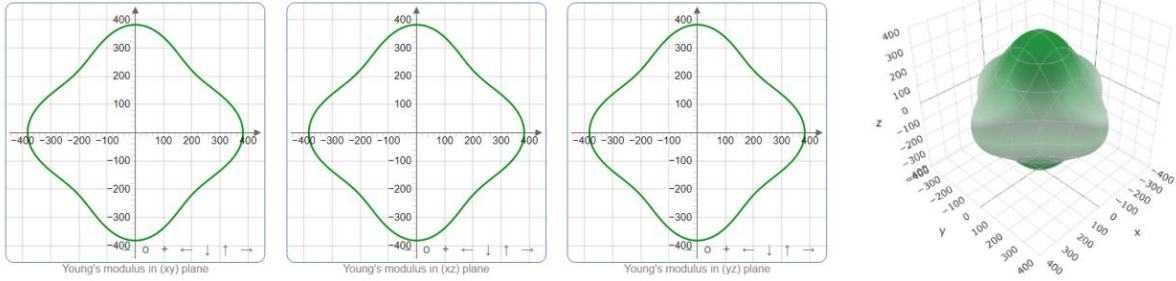
(b) Compressibility
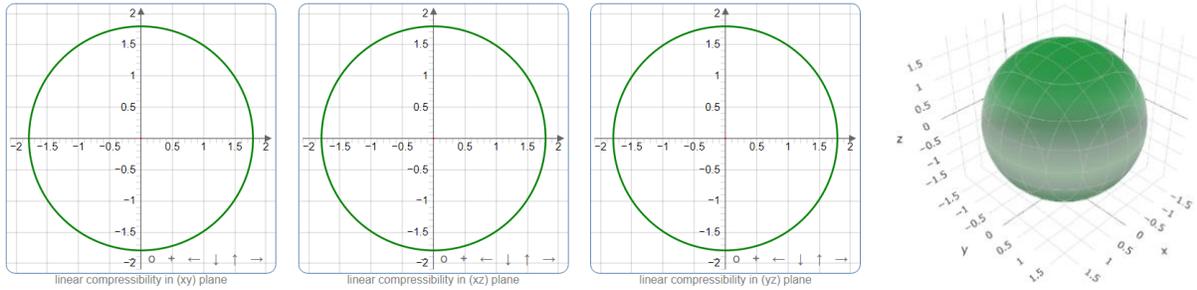
(c) Shear Modulus
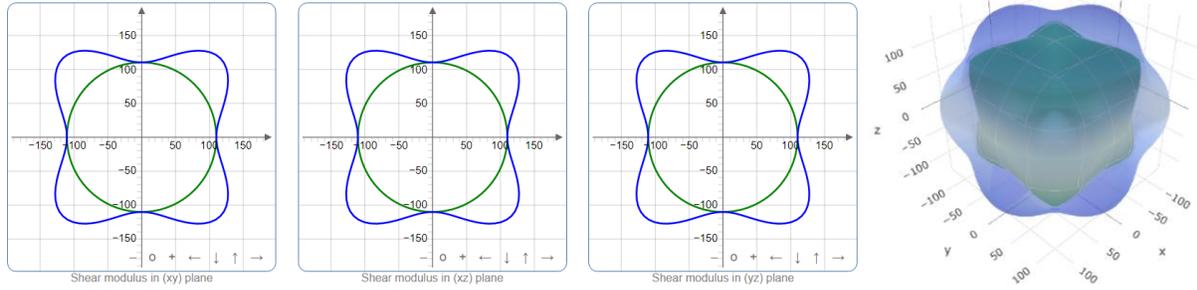
(d) Poisson's ratio
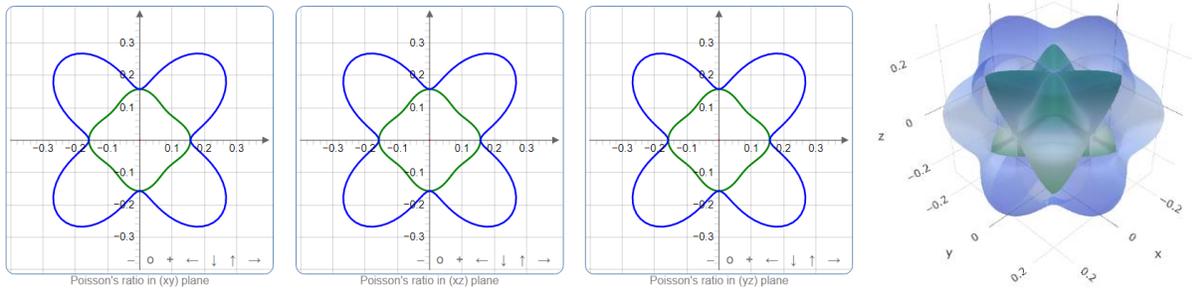

**Fig. S1:** The 2D and 3D plots of (a) $Y$, (b) $K$, (c) $G$ and (d) $v$ of RbTaO$_3$.

(a) Young's Modulus

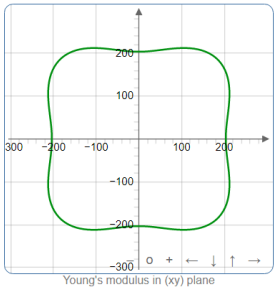 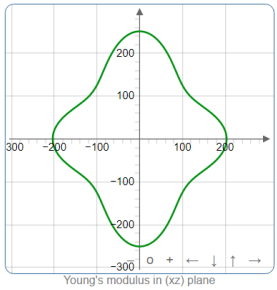 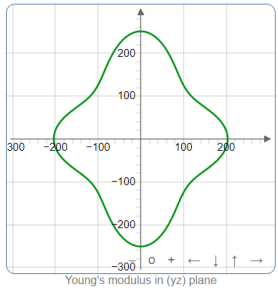 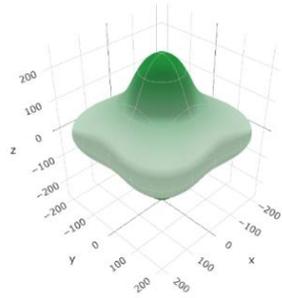

(b) Compressibility

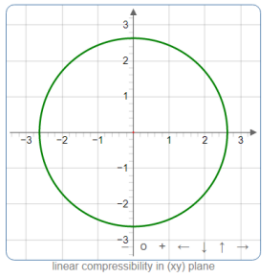 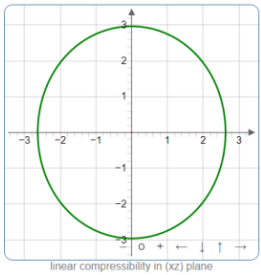 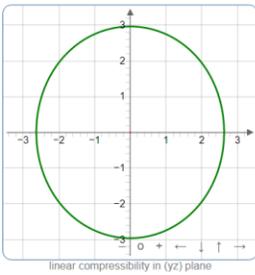 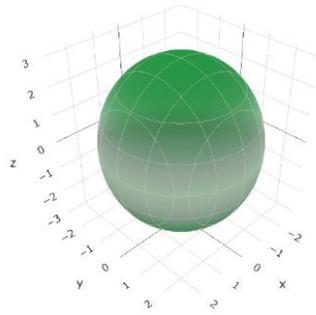

(c) Shear Modulus

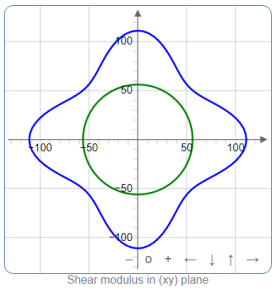 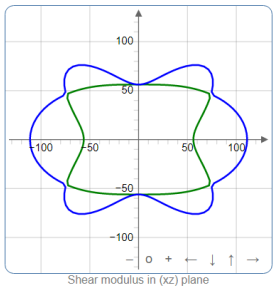 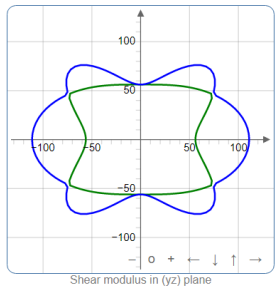 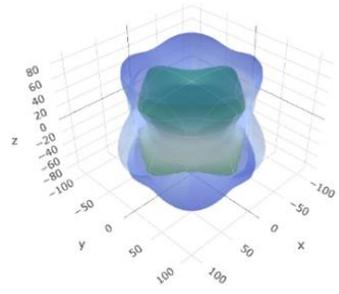

(d) Poisson's ratio

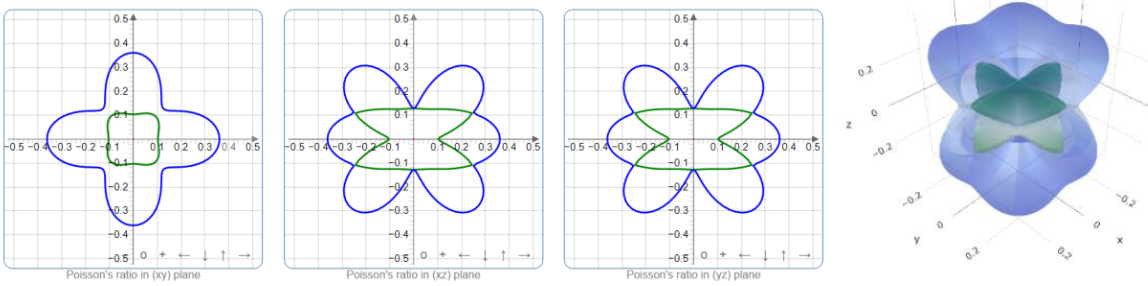

**Fig. S2:** The 2D and 3D plots of (a) $Y$, (b) $K$, (c) $G$ and (d) $\upsilon$ of RbTaO$_2$S.